\documentclass[conference]{IEEEtran}

\IEEEoverridecommandlockouts

\usepackage{tikz}
\usetikzlibrary{patterns, decorations.pathreplacing, calc, positioning,
  fit, arrows.meta, shapes.geometric}
\usepackage{algorithm}
\usepackage{algpseudocode}
\usepackage{booktabs}
\usepackage{multirow}
\usepackage{pgfplots}
\pgfplotsset{compat=1.18}
\usepackage{amsmath}
\usepackage{amssymb}
\usepackage{amsthm}
\usepackage{xcolor}
\usepackage[hidelinks]{hyperref}
\usepackage{cite}
\providecommand{\xmark}{\ensuremath{\times}}

\begin{document}

\title{Partitioned Tags, Shared Data: Reconciling Strict\\
       Cache Isolation with Write-Shared Coherence}

\author{\IEEEauthorblockN{Kartik Ramkrishnan, Stephen McCamant,
                          Antonia Zhai and Pen-Chung Yew}
        \IEEEauthorblockA{University of Minnesota\\
                          ramkr004@umn.edu}}

\maketitle

\begin{abstract}
Cache partitioning is among the strongest structural
defenses against eviction-based cache
side channels, yet a decade-old design issue has blocked
its widespread deployment in secure
shared-OS settings. The issue is that write-shared coherence
collapses under strict partitioning. We present
\textbf{SCP} (Secure and Coherent Partitioning),
which combines strict eviction isolation with
write-shared coherence by partitioning only the
tags, sharing a single data pool, and sizing
the data pool so
capacity-driven cross-partition eviction cannot
occur. Timing obfuscation
extends protections to the
inter-partition lookup path. 
Coherence-based leakage on shared-writeable lines is
mitigated by routing those writes
through to the LLC once a leakage threshold is crossed, which makes attacker
write probe latency independent of victim
activity. 

Using gem5 for implementation, SCP mitigates
Prime+Probe and Flush+Reload, which are the basis for more sophisticated cache attacks. We also demonstrate that a  
shared-writeable-line attack is mitigated. All these attacks yield results 
no better than random guessing. SCP's hardware  cost is a modest $+2.8\%$ LLC SRAM.
Performance matches DAWG within $0.3\%$ IPC on the  SPEC~CPU2017 benchmarks that we evaluated. Sharing-intensive microbenchmarks demonstrate a tunable security-performance tradeoff based on a system-specified leakage threshold.
\end{abstract}

\section{Introduction}
\label{sec:intro}

Modern processors rely on shared on-chip caches to
bridge the gap between fast cores and slow
memory~\cite{jouppi1990improving}, but those caches
have proven a rich source of microarchitectural
side-channels~\cite{ge2018survey,osvik2006cache,percival2005cache,bernstein2005cache,tromer2010efficient,liu2015last,gruss2016flush,disselkoen2017prime,briongos2020reload,yarom2014flush,oren2015spy,shusterman2019robust}.
A co-resident
attacker~\cite{ristenpart2009hey,inci2016cache,zhang2012cross}
can recover keys, break memory isolation, and amplify
transient-execution
attacks~\cite{kocher2019spectre,lipp2018meltdown,bulck2018foreshadow,schwarz2019zombieload,vanschaik2019ridl},
undermining cache-based isolation in confidential
computing~\cite{schluter2024heckler,schluter2024wesee,wilke2024tdxdown,schluter2025heracles,wiretap2025,meulemeester2025badram,vanschaik2024sgxfail,chiang2025reload}.

A long line of secure-cache work has converged on two
defense lineages. \emph{Randomization} obfuscates where a
victim's lines live in the cache, e.g.,  CEASER, CEASER-S,
ScatterCache, MIRAGE, INTERFACE, and their
successors~\cite{qureshi2018ceaser,qureshi2019new,werner2019scattercache,saileshwar2021mirage,kelemework2023interface,thoma2023scarf,unterluggauer2022chameleon,bhatla2024maya,bhatla2026avatar,bhatla2025sok}
make eviction-set construction exponentially harder
without forbidding sharing.
\emph{Partitioning} simply
denies sharing, e.g.,  PLcache, StealthMem, NoMo,
CATalyst, SecDCP, and DAWG carve the cache into
per-domain segments so that an attacker and a victim
never compete for the same
lines~\cite{wang2007new,kim2012stealthmem,intelcat,wang2016secdcp,kiriansky2018dawg}.
The two lineages offer different security versus performance tradeoffs, with partitioning enabling stricter security guarantees.  DAWG~\cite{kiriansky2018dawg} remains the canonical
strict-isolation partitioning baseline for defense against
contention-based and \textsc{Flush+Reload} attacks.

Table~\ref{tab:scp-intro-feature-matrix} summarizes
where this paper falls in that landscape along four
axes, namely, functionality \textbf{(F)}, security
\textbf{(S)}, coherence protocol performance \textbf{(P)}, and storage
\textbf{(O)}. 
Randomization
designs (rows 2--5) keep \textbf{(F)} but leak either
probabilistically and/or via the coherence channel and the heaviest variant pays 19.3\% storage overhead.
Partitioning (rows 6, 8--9) eliminates probabilistic leakages but may not support write-shared data \textbf{(F)}. 
The closest prior data coherent
defense, RAWS~\cite{ramkrishnan2024nonfusion} (row 7) is a hybrid of partitioning and randomization that keeps
\textbf{(F)} but pays a fixed per-miss 
delay. Secure and Coherent Partitioning's (SCP) distinctive feature is the
tunable \textbf{(S)} column. %
SCP lets the system specify where
on the performance$\leftrightarrow$security curve
to operate. SCP-WT (\emph{W}rite-\emph{T}hrough) is a high security configuration with no side-channels through the coherence protocol. SCP-P (\emph{P}ermissive) achieves a higher performance by allowing leakage up to a system specified threshold. The rest of the introduction explains why
 existing designs cannot reach this combination of features and what SCP changes structurally.

\begin{table}[t]
\centering
\scriptsize
\setlength{\tabcolsep}{4pt}
\renewcommand{\arraystretch}{1.1}
\caption{LLC-defense landscape on Functionality
(write-shared support), Security (leakage profile),
Performance, and storage Overhead. SCP is the only
design \emph{tunable} on (S) while keeping
(F)/(P)/(O).}
\label{tab:scp-intro-feature-matrix}
\begin{tabular}{l|cccc}
\toprule
Design & \textbf{(F)} & \textbf{(S)} & \textbf{(P)} & \textbf{(O)} \\
       & WR-share     & Leakage      & Coh. Perf.        & Storage \\
\midrule
\textsc{Baseline}                                            & $\checkmark$ & leaks      & $\checkmark$ & ---       \\
\textsc{CEASER-S}~\cite{qureshi2019new}                      & $\checkmark$ & coh.+prob. leaks       & $\checkmark$ & low       \\
\textsc{MIRAGE}~\cite{saileshwar2021mirage}                  & $\checkmark$ & coh.+prob. leaks & $\checkmark$  & $+19.3\%$ \\
\textsc{INTERFACE}~\cite{kelemework2023interface}            & $\checkmark$ & coh.+prob. leaks      & $\checkmark$ & $+7.4\%$  \\
\textsc{Avatar}~\cite{bhatla2026avatar}                      & $\checkmark$ & coh.+prob. leaks      & $\checkmark$ & $+1.5\%$  \\
\midrule
\textsc{DAWG}/\textsc{SecDCP}~\cite{kiriansky2018dawg,wang2016secdcp} & \xmark & no leaks & $\times$   & $+0.2\%$  \\
RAWS~\cite{ramkrishnan2024nonfusion}       & $\checkmark$ & coh.+prob. leaks      & \xmark       & high      \\
\textbf{\textsc{SCP-WT} (this work)}                            & \boldmath$\checkmark$ & \textbf{no leaks}    & \boldmath$\times$  & \boldmath$+2.8\%$ \\
\textbf{\textsc{SCP-P} (this work)}                            & \boldmath$\checkmark$ & \textbf{coh. leaks}    & \boldmath$\checkmark$  & \boldmath$+2.8\%$ \\
\bottomrule
\end{tabular}
\end{table}

Full partitioning has a structural problem that
randomization does not. It is incompatible with coherent
sharing of writeable lines. Consider two security
domains $A$ and $B$ that legitimately share a
writeable cache line, such as an OS spinlock, a
producer-consumer queue head, or any IPC
structure that crosses a trust boundary. Under
way-partitioning, $A$ and $B$ each hold the line
in their partitioned ways. There are now two cache
locations for the same physical line, and the
directory protocol has to keep them coherent. However, changing the behavior of the directory protocol, based on information in multiple partitions, would re-introduce information leakages.
To the best of our knowledge, existing directory
protocols offer no good solutions.

DAWG~\cite{kiriansky2018dawg} and its kin~\cite{saileshwar2021bce,dessouky2022chunked} handle this by either prohibiting
write-shared lines across partitions, restricting
sharing to read-only lines (where coherence reduces to
broadcast invalidation only on rare upgrades), or
accepting the cross-partition coherence traffic as a
known
limitation~\cite{kiriansky2018dawg,wang2016secdcp}.
None of these is satisfactory. Operating-system
synchronization, confidential-computing IPC, and any
shared-memory programming model that crosses the trust
boundary now hits a defense-policy wall that
less secure systems, e.g., randomization-based caches, do not face. %
MIRAGE~\cite{saileshwar2021mirage} introduces a useful
structural mechanism. It \emph{decouples} the tag
from the data array, connecting them with forward
and backward pointers. MIRAGE uses the decoupling
for set-associative-eviction freedom in
randomization. V-Way and
ZCache~\cite{qureshi2005vway,sanchez2010zcache}
use the same decoupling for demand-based
associativity and global replacement. Thus, tag/data
decoupling is a well-understood building block for security or performance purposes.

Inspired by the design features of tag/data decoupling and partitioning, \textbf{SCP} (\emph{Secure and Coherent
Partitioning}) is a partitioned cache architecture in which the
tag array is partitioned per security domain while the data
array is a single, unpartitioned, coherence-bearing
pool. Each tag entry contains a forward pointer into
the data array, and the coherence state of every line
is stored \emph{in the data entry}, not in any tag entry.

When a
domain misses in its tag partition, the controller
performs a parallel lookup across the other domains'
tag partitions. If the line is found, a new tag is
allocated in the missing domain's own partition,
 pointing at the existing data entry. The
data entry's reference count is incremented. No data is moved or duplicated.
Multiple tag entries in different partitions may
point at the same data entry, so a write-shared line
has exactly one coherence state and exactly one
writer-visible copy. Cross-partition coherence on
write-shared lines reduces to the standard
single-cache-line MSI/MESI transitions. 
The
multi-copy hazard that full partitioning creates
no longer exists.

Three security-relevant design features make this
work. First, tag miss responses are delayed to
memory-access latency before the data is returned,
so an attacker cannot distinguish a hit in
another partition from a true memory access.
Second, shared-writeable lines on eligible pages
exist only in shared (S) coherence state, once a system-specified leakage threshold is exceeded. Stores 
to these lines are written through to the LLC, so the E/M-versus-S
latency difference an attacker would otherwise
exploit is mitigated at its source. Third, the
data array is sized to match the total tag
count, so capacity-driven data eviction cannot
occur. A
data slot transitions from in-use to free only when its
reference count falls to zero
(\S\ref{sec:design:size}). No replacement policy
on the data array is needed, and no
cross-partition tag invalidation is ever induced
by another partition's tag-eviction pressure.

The key contributions of this work are summarised
below.

\begin{itemize}
\item \textbf{Architecture.} SCP applies
decoupled tag/data indirection to partitioning
rather than randomization, with a
sizing discipline that makes capacity-driven
data eviction structurally impossible. Per-domain
tag isolation and timing obfuscation close Prime+Probe and
Flush+Reload by construction. The remaining
timing leak on shared-writeable MSI/MESI lines
is closed by SCP-WT, which routes those stores
write-through to the LLC so that attacker
probe latency does not depend on victim  
activity. Alternatively, in SCP-P,  the write-through security feature is enabled when a system-specified leakage threshold is exceeded.
\item \textbf{Empirical attack-resilience.}
A gem5 prototype confirms that under SCP,
Prime+Probe, Flush+Reload and write-shared side-channels are mitigated. %
Ablation studies confirm that each mechanism is
independently necessary.
\item \textbf{Performance and hardware
feasibility.} %
On a SPEC~CPU2017 performance evaluation, SCP matches DAWG within
$0.3\%$ IPC on every completed benchmark. %
In line with performance expectations of cache partitioning, SCP outperforms
unpartitioned LRU by $0.5\%$ to $4.4\%$ on the
four mixed-pressure mixes. A counting Bloom
filter at the  front-end is projected
to save $82\%$ of cross-partition
tag-read energy at 16\,MiB. Stress tests on write-shared data microbenchmarks indicate up to 23\% performance overhead, when almost all accesses are writes on write-shared data.  %
\end{itemize}

\paragraph*{Paper organization}
\S\ref{sec:background} reviews the partitioning
and tag/data-decoupling lineages, \S\ref{sec:threat}
the threat model, \S\ref{sec:design} the SCP
architecture, \S\ref{sec:security} the security
analysis, \S\ref{sec:impl} the gem5 prototype,
\S\ref{sec:eval} the performance and security
evaluation, \S\ref{sec:related} related work.

\section{Background and Motivation}
\label{sec:background}

Cache side-channel defenses divide into two structural
families, namely,  \emph{randomization} (which hides where a
line lives) and \emph{partitioning} (which forbids
two domains from sharing a line). SCP belongs to the
partitioning family but borrows a structural
mechanism, i.e., decoupled tag/data indirection, from the
randomization side. This section reviews both
lineages and the unsolved coherence problem of strict
partitioning.

\subsection{Cache attacks the defenses target}
\label{sec:bg:attacks}

\paragraph*{Contention-based attacks}
\textsc{Prime+Probe}~\cite{osvik2006cache,liu2015last,morgan2025sliceslicebaby,kessous2024prune,zhao2024llcfeasible}
fills a target cache set with attacker lines, lets the
victim run, and times re-accesses; long re-access
latencies indicate that the victim evicted attacker
lines from the set, leaking the victim's set
footprint. Variants such as \textsc{Evict+Time},
\textsc{Evict+Reload}, and
\textsc{Reload+Refresh}~\cite{briongos2020reload,disselkoen2017prime} differ
in measurement primitive but share the requirement
that the attacker can place lines in sets the victim
also uses.

\paragraph*{Address-based attacks}
\textsc{Flush+Reload} and
\textsc{Flush+Flush}~\cite{yarom2014flush,gruss2016flush}
require shared memory. The attacker flushes a known
shared line, lets the victim run, and times
re-loads. A fast re-load betrays a victim access. The
defense is to prevent the attacker from caching the
\emph{same physical line} as the victim or to deny
\texttt{clflush} on cross-domain lines.

\paragraph*{Occupancy and coherence channels}
Cache-occupancy
attacks~\cite{shusterman2019robust,chakraborty2025occupancy,cao2025mirage}
read coarse aggregate residency rather than
set-level eviction, sidestepping randomization.
\emph{Coherence-induced} channels exploit invalidation
or upgrade traffic on shared writeable
lines~\cite{purnal2021prime,zhang2024invalidate}. A 
writer's coherence broadcast is observable as a
slowdown on a sharer's subsequent access.

\subsection{Tag/Data Decoupling in Randomization}
\label{sec:bg:rand}

MIRAGE~\cite{saileshwar2021mirage} took the further
step of \emph{decoupling} tags and data: the tag array
is over-provisioned (e.g., $1.75\times$ the number of
data slots), each tag entry holds a forward pointer
to a data slot, and each data slot a reverse pointer
back. The original motivation was to absorb the
\emph{spill tail} of randomized placement under
balls-into-bins analysis. INTERFACE~\cite{kelemework2023interface}
refined the same decoupling with a $2{+}2$-skew
arrangement at lower storage overhead. The structural
shape is a tag array indirection-pointed at a separately
addressed data array. This is what SCP reuses.

V-Way~\cite{qureshi2005vway} and
ZCache~\cite{sanchez2010zcache} predate the
randomization line and used the same decoupling for
demand-based associativity and global replacement,
respectively. The decoupled tag/data idea is therefore
a generic building block. A key novelty in SCP is the coupling to a partitioning
security model.

\subsection{Partitioning caches and the coherence
problem}
\label{sec:bg:part}

Partitioning works by denying co-residency. PLcache
(line-pinning)~\cite{wang2007new}, page coloring, and
StealthMem~\cite{kim2012stealthmem}, Intel
CAT~\cite{intelcat}, CATalyst, NoMo, and
DAWG~\cite{kiriansky2018dawg} all carve the cache
into per-domain subsets so that an attacker simply
cannot place a line in a victim's set.
SecDCP~\cite{wang2016secdcp} adds asymmetric
\emph{one-way} dynamic sizing so the partition's size
itself is not a side-channel.

\paragraph*{The coherence problem}
The hidden cost of strict partitioning is its
interaction with coherence on \emph{write-shared}
lines. A line that two domains both legitimately
write, e.g., an OS lock variable, a confidential-computing
IPC ring buffer head, or a shared work-stealing
deque, must, under way-partitioning, occupy a way in
each domain's partition. There are then two physical
storage locations for one logical line. The coherence
protocol now has three unattractive options:
\begin{enumerate}
\item \textbf{Hold both copies and keep them coherent
in the LLC.} Every $A$-write triggers a $B$-side
invalidation; $B$'s subsequent re-read misses to
memory (or, in inclusive hierarchies, to the LLC
copy). The cross-partition coherence traffic is itself
observable as timing on $B$'s next access, recreating
the side-channel partitioning was meant to close.
\item \textbf{Pin the line to one partition.} The
non-owning partition cannot cache it and pays
LLC-bypass latency on every access. Useful only for
mostly-read-only data.
\item \textbf{Forbid cross-partition write-sharing
entirely.} The simplest and the policy DAWG-style
deployments effectively
adopt~\cite{kiriansky2018dawg}, but it is a real
usability. It blocks shared-memory IPC
across security boundaries, OS-driven cross-domain
coordination, and any programming model that needs
write-shared structures across the partition.
\end{enumerate}

The randomization line does not have this problem
because it never partitioned in the first place. A 
shared writeable line lives in exactly one slot, and
coherence is single-copy MSI/MESI. The cost the
randomization line pays instead is large tag-storage
overhead and probabilistic, not deterministic,
isolation. SCP is the architecture that takes the
randomization line's single-copy data-array structure
\emph{without} its randomization, and pairs it with
deterministic tag-array partitioning.

\subsection{Moving coherence metadata to the data array}
\label{sec:bg:coherence}

An important architectural move that makes SCP work is to
locate the canonical coherence state of each line
\emph{in the data entry}, not in any tag entry. In a
conventional cache, the tag holds \{address, valid,
state, dirty, LRU\}. Under
partitioning, this means each partition's per-line
state is local to that partition, so the protocol
must reconcile multiple states across partitions on
every coherence event. This is the source of a data consistency correctness hazard for programs using shared data across domains.

Under SCP, the data entry holds \{state, dirty,
refcount\}. The tags hold \{address, valid,
forward pointer\}. The per-line sharer vector
that the coherent LLC already maintains is
reused for any cross-domain invalidate;
SCP adds no per-data-slot owner mask. There is exactly one coherence
state per line because there is exactly one data
entry per line, regardless of how many tag entries
across how many partitions point at it. Every
coherence transition (S$\to$M, M$\to$I, etc.) operates
on a single state machine. %
The next two sections
make this precise.

\section{Threat Model}
\label{sec:threat}

We adopt a standard secure-cache threat
model~\cite{kiriansky2018dawg,saileshwar2021mirage,wang2016secdcp,bhatla2025sok}.
The attacker (domain $A$) and victim (domain $B$)
share a last-level cache, either via SMT on the same
physical core or as separate cores in a CMP. Each
domain runs its own user-mode code on a commodity
ISA. The attacker can:
\begin{itemize}
\item issue arbitrary memory accesses to its own
address space;
\item time those accesses with cycle-accurate
counters (e.g., \texttt{rdtscp});
\item use \texttt{clflush} on lines mapped into its
own address space;
\item construct eviction sets via standard
techniques~\cite{liu2015last,vila2019theory,purnal2021systematic};
\end{itemize}

\paragraph*{Adversary goal}
Recover a function of the victim's secret-dependent
memory access pattern. A canonical example is AES
T-table indices~\cite{osvik2006cache,tromer2010efficient}.
We do not assume secret-dependent control flow is
available to the attacker through other channels;
this paper concerns the cache channel specifically.

\paragraph*{ Eliminating the side
channels} Cache partitioning has, since DAWG and
its predecessors, had one structural target, namely, to deny
the attacker the ability to evict victim lines or
to be evicted by them. SCP achieves this target
in full and structurally:
\begin{itemize}
\item \textbf{Contention-based attacks.}
  Attacker tags live in $A$'s tag partition,
  victim tags in $B$'s, and the two never share a
  tag set or way. $A$'s eviction set populates
  only $A$-tags. The victim accesses cannot evict
  them, and the
  contention-based primitive
  extinguishes (\S\ref{sec:sec:contention}).
\item \textbf{\textsc{Flush+Reload} on
  non-shared lines.} An attacker reload of a
  line not in $A$'s tag partition escalates to
  \textsc{PeerProbe}, which returns at a cache miss timing whether or not the
  victim has the line cached, removing the
  hit/miss timing distinction
  (\S\ref{sec:sec:contention}).
\item \textbf{Cross-partition tag invalidation
  via data-array eviction.} This channel is closed by
  construction. Capacity-driven data eviction
  does not happen
  (\S\ref{sec:design:size}). Tag eviction in
  partition $d$ never reaches another partition's
  tag array (\S\ref{sec:sec:contention}).
\item \textbf{Aggregate-occupancy attacks on
  non-shared workloads.} Each partition's
  observable footprint depends only on its own
  accesses plus the shared-line subset, so on
  workloads with no genuinely shared lines the
  occupancy channel is closed
  (\S\ref{sec:sec:occupancy}).
\end{itemize}
The eviction-based side-channel that motivated a
decade of secure-cache work is therefore
\emph{structurally absent}, not bounded
probabilistically. This is the central
positive claim of the paper.

\paragraph*{The inherent coherent
shared-line channel} If $A$ and $B$ deliberately
share a writeable cache line, e.g., a
cross-domain spinlock or IPC ring head, the
coherence protocol must invalidate one when the
other writes. This is information flow that
any cache architecture supporting
coherent cross-domain writeable sharing
necessarily transmits, including the
unpartitioned baseline LRU
cache~\cite{yarom2014flush,gruss2016flush}. SCP
does not introduce a new channel here. Furthermore, the write, by definition,  necessarily leaks information by updating the contents of write-shared data. %
The point of SCP, relative to DAWG, is that
this is now the \emph{only} cross-domain channel.
DAWG has the same channel \emph{plus} forced
incoherence or LLC-bypass on every write-shared
line, which SCP eliminates by construction
(\S\ref{sec:design}).

\subsection{Controlling  Leakage.}
\label{sec:design:eligibility}
SCP adds one TLB-attached field per page, which
the OS sets at page allocation
time. The field encodes a $2$-bit mode selecting
how the page handles cross-domain
shared-writeable lines. Three possible modes are  \texttt{SCP-adaptive}
(default), \texttt{SCP-WT}, or
\texttt{SCP}. %

\paragraph*{Three  modes}
\label{sec:design:wt}
\noindent\emph{(i)~\texttt{SCP-WT}.} L1 and
L2 are forced write-through on these lines, and
stores do not produce a private dirty copy. The
OS selects this through the page's memory type,
alongside the standard write-back,
write-combining, and uncacheable choices.
Standard store-merging hardware in the core
absorbs adjacent stores within a cache line. The
E/M-versus-S timing difference an attacker would
otherwise measure does not exist on these lines,
so probe latency is victim-independent under any
attacker rate.
\noindent\emph{(ii)~\texttt{SCP-adaptive}}
(default for new shared-writeable pages). The
line behaves as a normal MSI/MESI line until the LLC
counts more than $T_{\text{leak}}$ victim-driven
E/M$\to$S transitions on the page within a
specified time window (e.g., $1$\,ms), at which point the page is
auto-promoted to \texttt{SCP-WT}. %
Pre-promotion leakage is
bounded by $T_{\text{leak}}$ events per window.
Post-promotion the page is structurally closed.
With $T_{\text{leak}}\!=\!0$ this collapses to
always-WT. With $T_{\text{leak}}\!=\!\infty$ this collapses to
stock MESI/MSI. The prototype default is
$T_{\text{leak}}\!=\!100$ per $1$\,ms. Hardware
cost is one per-page counter near the TLB. Pages that never trip the budget
pay nothing.
\noindent\emph{(iii)~\texttt{SCP}.}
The page runs base SCP with full tag isolation
against eviction-based attacks
(\textsc{Prime+Probe}/\textsc{Flush+Reload}
remain closed by construction), unrestricted
cross-domain shared-writeable access, and the
M$\to$S coherence channel left open at zero
runtime overhead. The right choice for pages
where the cross-domain timing channel on those
lines is not part of the threat (e.g., shared
lookup tables where the access patterns are not secret). %

\paragraph*{Out of scope}
Out-of-cache microarchitectural channels (memory
bus, DRAM rowbuffer, ring-bus, TLB)
\cite{rodrigues2024interconnectsurvey,rodrigues2024busted,vanoverloop2025tlblur},
transient-execution channels
\cite{kocher2019spectre,lipp2018meltdown,bulck2018foreshadow,schwarz2019zombieload},
and hypervisor or platform-firmware compromise
are orthogonal.
SCP does not  preclude the use of other defensive strategies.

\section{The SCP Architecture}
\label{sec:design}

This section presents the SCP cache structure
(\S\ref{sec:design:struct}), the operation paths for
hit, find, miss, write, and eviction
(\S\ref{sec:design:ops}), the data-array sizing
that makes capacity-driven eviction structurally
impossible (\S\ref{sec:design:size}), and the
constant-time miss-latency mask
(\S\ref{sec:design:mask}).
Section~\ref{sec:design:invariants} consolidates the
invariants the security argument relies on.

\subsection{Structure}
\label{sec:design:struct}

\begin{figure}[t]
\centering
\resizebox{\columnwidth}{!}{%
\begin{tikzpicture}[
  font=\scriptsize,
  every node/.style={inner sep=1.5pt},
  tagpart/.style={draw, rounded corners=2pt,
    minimum width=2.7cm, minimum height=0.45cm,
    align=center},
  datalbl/.style={draw, rounded corners=2pt,
    minimum width=2.9cm, minimum height=0.45cm,
    align=center, fill=blue!8},
  bf/.style={draw, dashed, rounded corners=2pt,
    minimum width=1.0cm, minimum height=0.35cm,
    align=center, fill=yellow!15, font=\tiny},
  arr/.style={-{Latex[length=1.5mm]}, thin},
  cfarr/.style={-{Latex[length=1.5mm]}, thin, dashed,
    purple!70!black}
]
\node[tagpart, fill=red!10]   (tA1) at (0,2.2) {A.tag\,[0]: addr,$\,$fp};
\node[tagpart, fill=red!10]   (tA2) at (0,1.7) {A.tag\,[1]: addr,$\,$fp};
\node[tagpart, fill=red!10]   (tA3) at (0,1.2) {A.tag\,[$W_A$$-$1]: $\ldots$};
\node[fit=(tA1)(tA3), draw, dashed, rounded corners=2pt,
      label={[font=\scriptsize\bfseries]above:Domain A tag partition}]
      (boxA) {};

\node[tagpart, fill=green!10] (tB1) at (0,0.0)  {B.tag\,[0]: addr,$\,$fp};
\node[tagpart, fill=green!10] (tB2) at (0,-0.5) {B.tag\,[1]: addr,$\,$fp};
\node[tagpart, fill=green!10] (tB3) at (0,-1.0) {B.tag\,[$W_B$$-$1]: $\ldots$};
\node[fit=(tB1)(tB3), draw, dashed, rounded corners=2pt,
      label={[font=\scriptsize\bfseries]below:Domain B tag partition}]
      (boxB) {};

\node[datalbl] (d0) at (5.2,2.0)  {data\,[0]: state,$\,$rc};
\node[datalbl] (d1) at (5.2,1.5)  {data\,[1]: state,$\,$rc};
\node[datalbl] (d2) at (5.2,1.0)  {data\,[2]: state,$\,$rc};
\node[datalbl] (d3) at (5.2,0.5)  {data\,[3]: state,$\,$rc};
\node[datalbl] (d4) at (5.2,0.0)  {data\,[4]: state,$\,$rc};
\node[datalbl] (d5) at (5.2,-0.5) {$\ldots$};
\node[datalbl] (d6) at (5.2,-1.0) {data\,[$N$$-$1]: state,$\,$rc};
\node[fit=(d0)(d6), draw, rounded corners=2pt,
      label={[font=\scriptsize\bfseries]above:Shared data array
        (one coherence state per line)}] (boxD) {};

\node[bf] (bf) at (-2.4, 0.6) {Bloom\\filter\\$m{,}K$};

\draw[arr, red!70!black]
      (tA1.east) -- node[above, sloped, font=\tiny\bfseries,
                          color=red!70!black, pos=0.55] {fwd ptr}
      (d0.west);
\draw[arr, red!70!black]   (tA2.east) -- (d2.west);
\draw[arr, red!70!black]   (tA3.east) -- (d4.west);
\draw[arr, green!50!black] (tB1.east) -- (d0.west);  %
\draw[arr, green!50!black] (tB2.east) -- (d3.west);
\draw[arr, green!50!black] (tB3.east) -- (d4.west);  %

\draw[cfarr] (bf.east) -- node[above, font=\tiny, sloped]
      {if BF says ``maybe''} (boxA.west);
\draw[cfarr] (bf.east) -- (boxB.west);
\end{tikzpicture}}

\caption{SCP architecture. \emph{Tag arrays}
(left, red and green) are way-partitioned per
security domain. The \emph{data array} (right,
blue) is a single shared pool whose entries hold
the coherence state and reference count.
Multiple tags in different partitions may point
at the same data entry, giving exactly one
coherence state per cache line. A counting
\emph{Bloom filter} (yellow) sits at the
\textsc{PeerProbe} front-end. It is consulted
before the per-partition tag-array scan,
short-circuiting on negative.}
\label{fig:scp-arch}
\end{figure}
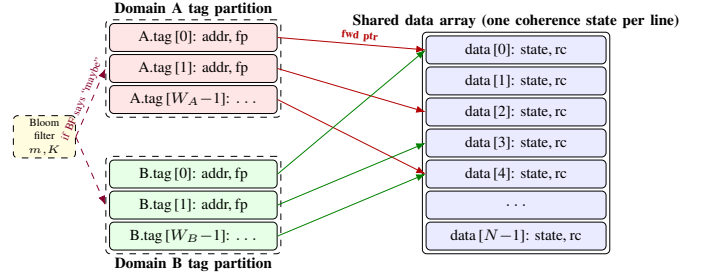

Figure~\ref{fig:scp-arch} shows the SCP cache architecture. The
cache consists of three components.

\paragraph*{(1) Per-domain tag partitions.} For
$D$ security domains, the tag array is partitioned
into $D$ disjoint regions. Each domain $d$ owns
$W_d$ tag ways per set (with $\sum_d W_d = W$, the
physical tag-array associativity). Each tag entry
holds:
\begin{center}
\begin{tabular}{l@{\,\,}l}
\texttt{addr} & line address \\
\texttt{valid} & validity bit \\
\texttt{fp} & forward pointer (\,$\log_2 N$\,bits) \\
\texttt{lru} & per-domain LRU bits \\
\end{tabular}
\end{center}
where $N$ is the number of data-array entries.
Critically, the tag does \emph{not} hold the
coherence state, the dirty bit, or the data. Those
instead live with the data entry.

\paragraph*{(2) Shared data array} A single pool of
$N$ data entries, addressed by a flat index from
$0$ to $N-1$. Each entry holds:
\begin{center}
\begin{tabular}{@{}l@{\,\,}l@{}}
\texttt{state}    & MSI/MESI state (M/E/S/I) \\
\texttt{dirty}    & dirty bit (M-state) \\
\texttt{refcount} & number of partitions pointing here \\
\texttt{data}     & cache-line data \\
\end{tabular}
\end{center}
The line address is held only in the tag entries
that point at this data slot; the data entry has
no redundant address copy. SCP does not introduce
a per-data-slot \emph{owner mask} either: any
broadcast invalidate on the line uses the sharer
vector that a coherent LLC already maintains
for the line, which already encodes which
domains hold the line.

\paragraph*{(3) Cross-partition find} A
parallel tag-lookup datapath that, on a miss in
partition $d$, broadcasts an address-match request
to the other $D-1$ tag partitions in parallel. The
result is either a single hit (find in some
partition $d' \neq d$) or no hit (true miss). The
find operation is timing-equalized to memory-miss
latency before returning data, by the mask described
in \S\ref{sec:design:mask}. Figure~\ref{fig:scp-arch}
shows the structure for $D{=}4$ explicitly. The
forward pointers, the data-side coherence state,
and the mask on the invalidate-broadcast path
are the three primitives the rest of the
architecture composes.

\paragraph*{Storage cost}
Relative to a baseline cache with $S$ data slots
and $S$ tag entries (one per data slot), SCP's
storage overhead has two components:
(i)~the data entry's extra fields (state,
refcount), about $2{+}\lceil\log_2(D{+}1)\rceil$
bits per data entry. The existing per-line sharer vector of a
coherent LLC already encodes which domains hold
the line (ii)~the forward pointer in each tag
entry, $\log_2 N$ bits. For $D{=}4$, $40$-bit
PA, $16$-MiB LLC, $N\!=\!S\!=\!2^{18}$, this
evaluates to $+2.6\%$ LLC SRAM
(\S\ref{sec:eval:storage}, Tab.~\ref{tab:scp-storage}),
on the same order as DAWG and far below
MIRAGE-class designs.

\subsection{Operations}
\label{sec:design:ops}

We describe five elementary operations. They are 
\textsc{Lookup}, \textsc{PeerProbe}, \textsc{Allocate},
\textsc{Write}, and \textsc{Evict}. Algorithm pseudocode
appears in Appendix~\ref{app:scp-algos}.

\paragraph*{Lookup of address $X$ by domain $d$}
The controller reads tag entries for the appropriate
set in $d$'s tag partition. If a valid tag with
\texttt{addr}$\,=\,$\!$X$ exists, it dereferences the
forward pointer to the data entry and reads the
state. The state must be M, E, or S to deliver a
hit %
If no such tag exists, the request
escalates to \textsc{PeerProbe}.

\paragraph*{Cross-partition find} The controller
issues a parallel address-match (PeerProbe) against
the other $D-1$ tag partitions in $d$'s LLC slice.
Algorithm~\ref{alg:peerprobe-body} summarises the
path. If the line is found in partition $d'$, the
controller allocates a new tag entry in $d$'s
partition (evicting an LRU $d$-tag if necessary),
sets the new tag's forward pointer to the shared
data entry, sets bit $d$ in the data entry's owner
mask, and increments \texttt{refcount}. The data
entry's coherence state is updated. %
The latency-mask logic delays the response so the
total elapsed time matches a memory-miss latency
(\S\ref{sec:design:mask}). If no other partition
holds the line, the request proceeds to
\textsc{Allocate}.

\begin{algorithm}[h]
\small
\caption{SCP \textsc{PeerProbe} on tag-partition miss.}
\label{alg:peerprobe-body}
\begin{algorithmic}[1]
\Function{PeerProbe}{$X, d$}
  \State $E \gets \emptyset$
  \For{$d' \in \mathit{Domains} \setminus \{d\}$ \textbf{in parallel}}
    \For{tag $t$ in the addressed set of $\Call{TagPartition}{d'}$}
      \If{$t$.\texttt{valid} $\wedge$ $t$.\texttt{addr}$\,=\,$$X$}
        \State $E \gets E \cup \{$data array$[t.\texttt{fp}]\}$
      \EndIf
    \EndFor
  \EndFor
  \If{$|E| > 0$}
    \State pick $e \in E$ \Comment{at most one if invariants hold}
    \State allocate new tag $t_d$ in $\Call{TagPartition}{d}$
    \State $t_d$.\texttt{addr} $\gets X$;\;
           $t_d$.\texttt{fp} $\gets$ id($e$)
    \State $e$.\texttt{refcount} $\gets e$.\texttt{refcount} $+ 1$
    \State delay response to $T_{\text{miss}}$
    \State \Return \textsc{HitFind}, $e$
  \Else
    \State delay response to $T_{\text{miss}}$
    \State \Return \Call{Allocate}{$X, d$}
  \EndIf
\EndFunction
\end{algorithmic}
\end{algorithm}

\paragraph*{Allocate (true miss).} The controller
issues a memory request for $X$. While memory is
in flight, it selects a tag-array victim from
$d$'s tag partition (per-partition LRU). The
victim tag's eviction (\textsc{Evict}~below)
decrements its referenced data entry's refcount.
If that refcount reaches zero, the corresponding
data slot becomes free and is the natural target
for the new line. The data array's sizing
(\S\ref{sec:design:size}) guarantees a free slot
is available. On memory return, the freed data
slot is repurposed.
The new tag in $d$'s partition forward-points at
it.

\paragraph*{Write to address $X$ by domain $d$}
Same as \textsc{Lookup} except the access is a write.
On a hit:
\begin{itemize}
\item if the data state is M (the writer already
  owns it exclusively), update the line and remain in
  M;
\item if the state is E (this writer's partition is
  the sole holder), transition to M and update;
\item if the state is S (other partitions have tags
  pointing here too), trigger an
  \textsc{S$\to$M Upgrade}. %
\end{itemize}

\paragraph*{Evict} Tag eviction is the \emph{only}
eviction event SCP triggers. When partition $d$
removes a $d$-tag (e.g., to make room for a new
$d$-tag in the same set under per-domain LRU), the
controller clears partition $d$'s bit in the
LLC's per-line sharer vector and decrements the
data entry's refcount.
\begin{itemize}
\item If \texttt{refcount} remains $> 0$, the data
  entry stays valid. At least one tag in some other
  partition still points at it. Nothing else
  happens. \textbf{No tag in any other partition is
  touched.}
\item If \texttt{refcount} reaches $0$, no tag in
  the cache references this data entry any longer.
  The controller transitions the entry's coherence
  state to I, writes back the data if it was M or
  dirty, and the slot is now free for the next
  allocate.
\end{itemize}
This is the architectural payoff of decoupling. A 
tag eviction in partition $d$ touches $d$'s tag
array and the shared data entry's refcount, and
\emph{nothing else}. There is no replacement
policy on the data array, no candidate-victim
selection, and no propagation of $d$'s capacity
pressure into any other partition's tag state.

\subsection{Data-array sizing to eliminate capacity-driven
eviction}
\label{sec:design:size}

The shared data array is sized so that capacity
pressure on the data side is structurally
impossible. Let $W = \sum_d W_d$ be the total
tag-array associativity (summed over the $D$
per-domain partitions) and $S$ the number of sets.
Total tag entries are $N_{\text{tag}} = W \cdot S$.
We size the data array $N = N_{\text{tag}}$.

\begin{quote}\itshape
\textbf{Sizing invariant.} The number of data slots
equals the total number of tag entries.
\end{quote}

\paragraph*{Why this works} In a worst case for
sharing, e.g., no two domains share any line, each tag
holds a unique forward pointer to a distinct data
entry. Total live data entries are bounded by total
valid tags, which is bounded by $N_{\text{tag}} =
N$. The data array is large enough by construction.

In a case that exercises sharing, e.g., multiple partitions
hold tags pointing at the same line, several tags
forward-point at one data entry, the live data
count drops below $N$, and the spare data slots are
simply unused. The capacity is paid once, but
duplicated under sharing it serves more partitions
than DAWG's per-partition data quota would.

\paragraph*{No replacement policy on
the data array.} Because no allocate is ever
blocked by lack of space, the data array needs no
replacement decision at all. Data slots transition
between two states: \emph{live} (refcount $> 0$,
holding the canonical M/E/S/I state of the line)
and \emph{free} (refcount $= 0$, available for the
next allocate). The transition live$\to$free is
caused only by tag eviction reducing refcount to
zero.

\paragraph*{No cross-partition
tag-invalidation back-channel} Tag eviction in
partition $d$ touches $d$'s tag array and the
referenced data entry's refcount, period. No other
partition's tag array can be invalidated by $d$'s
capacity pressure, because no data entry is
``replaced''. A data slot only frees when every
tag pointing at it has already been evicted by its
own partition's policy. The eviction-based
channel that partitioning was designed to
eliminate is fully closed by the
\emph{decoupling}, without any need for an
ownership-aware replacement policy.

\subsection{ Miss-latency obfuscation}
\label{sec:design:mask}

A \textsc{PeerProbe} that succeeds (line found in
another partition) returns data faster than a true
memory miss. If exposed, this latency difference
would let an attacker distinguish ``victim has this
line cached'' from ``victim does not.'' SCP closes
this distinction with a fixed-latency timing obfuscation. Every
\textsc{PeerProbe} (success or failure) returns
data only after $T_{\text{miss}}$ cycles, where
$T_{\text{miss}}$ is a memory miss latency.
Successful finds buffer the data and stall
delivery to match.

\paragraph*{Cost of the timing obfuscation} The mask makes a
cross-partition shared line behave, from the
attacker's perspective, like a true memory miss.
The performance benefit of cross-partition
sharing is then \emph{indirect}. The next access to
the same line by domain $d$ hits the now-allocated
$d$-tag at $L1$/$L2$ latency. Thus, the steady-state
benefit on shared workloads remains. Only the first
\textsc{PeerProbe} per shared line per partition
pays the mask. %

\subsection{Coherence protocol independence}
\label{sec:design:protocol}

SCP's tag-isolation and refcount machinery are
protocol-independent. For instance, the same surface operates
over MESI, MSI, or MI. 
The
mechanism is also transport-agnostic. Directory or snoop based protocols can be used. %

\subsection{Design invariants}
\label{sec:design:invariants}

The remainder of the paper relies on four
invariants the architecture maintains:

\begin{enumerate}
\item \textbf{(Tag isolation)} Domain $d$'s
  accesses, namely, hits, misses, and tag evictions, modify
  the state of $d$'s tag partition only. No other
  partition's tag array is ever modified as a side
  effect of $d$'s accesses, except via an explicit
  \textsc{S$\to$M Upgrade}~on a genuinely shared
  writeable line.

\item \textbf{(Single coherence state)} Every cache
  line resides in exactly one data entry. Its
  M/E/S/I state is held by that data entry. There
  is no per-tag duplicate state to synchronize.

\item \textbf{(Timing Obfuscation)} Every
  \textsc{PeerProbe}, whether success or failure, appears as a cache miss from the requesting
  domain's timing perspective.

\item \textbf{(Refcount conservation)} For every
  data entry, \texttt{refcount}  equals the
  number of valid tags across all partitions whose
  forward pointer targets the entry. A data entry
  is \emph{free} iff \texttt{refcount}$\,=\,0$.
  Capacity-driven data evictions never occur.
  \item \textbf{(Leakage Limit)}
  $E/M \rightarrow S$ downgrades occur after a system-specified number write-based side-channels transmissions occur.
\end{enumerate}

\section{Security Analysis}
\label{sec:security}

SCP structurally eliminates the eviction-based
side-channel and reduces the residual cross-domain
signal to the inherent write-based information flow.
Under the SCP invariants, every non-write cross-domain flow
falls into one of three categories.
(\emph{a})~eliminated by tag isolation,
(\emph{b})~eliminated by the timing obfuscation, or
(\emph{c})~eliminates write-based side-channel measurement. We treat each attack class
in turn.

\subsection{Eviction- and flush-based attacks}
\label{sec:sec:contention}

\textsc{Prime+Probe}, \textsc{Evict+Time}, and
\textsc{Reload+Refresh} all rely on the attacker
constructing an eviction set whose lines map to
the same set as the victim's target line. The
attack then uses victim-induced eviction as a
timing signal: the attacker primes the set with
its own lines, lets the victim run, and times a
re-access to detect which of its primed lines
have been evicted. Under SCP, invariant~1 (tag
isolation) confines an attacker's tag entries to
its own partition. The victim's tag entries are
in a disjoint partition. The two partitions
never share a tag set or way, so the attacker's
``eviction set'' against the victim does not
exist. Nothing the attacker can do at the tag
layer evicts the victim's lines. Data-array
pressure cannot reach the victim either, because
the $N{=}S$ sizing of \S\ref{sec:design:size}
makes capacity-driven data eviction structurally
absent. \textsc{Flush+Reload} on a line $X$ that
is not in $A$'s tag partition forces $A$'s
reload to escalate to PeerProbe. The PeerProbe
mask returns at memory-miss latency
$T_{\text{miss}}$ whether or not the victim is
caching $X$ in another partition, so the
hit-versus-miss timing distinction the attack
relies on is gone. The empirical evidence of
this collapse is in Table~\ref{tab:scp-cycle-uniform}
in \S\ref{sec:eval:sec}: $v{=}1$ and $v{=}0$
attacker-probe latencies agree on the gem5 sweep.

\subsection{Coherent shared lines}
\label{sec:sec:coherence}

When two tags from different partitions point at
the same data entry, an MSI/MESI write by either
partition triggers a coherence transaction. The
latency of that transaction depends on whether a
peer holds the line. A write to a line in M
completes locally with no bus traffic, while a
write to a line in S incurs an
$\textsc{S}\!\to\!\textsc{M}$ upgrade
(invalidating other holders) before completing.
The same observable channel exists under MESI's
E state. A victim line in E that is read by a
peer downgrades $\textsc{E}\!\to\!\textsc{S}$,
so the victim's next write needs an
$\textsc{S}\!\to\!\textsc{M}$ upgrade where it
would otherwise have silently upgraded
$\textsc{E}\!\to\!\textsc{M}$. On a conventional
cache, that M/E-versus-S latency difference is
what a co-resident attacker would measure to
learn whether a peer domain is currently
caching the same line. SCP closes both
transitions by routing shared-line stores
\emph{write-through} to the LLC for pages
flagged eligible-WT. Every store on a
write-through line follows the same path
regardless of peer state, so the line never
sits in M or E long enough to be observed. Attacker probes on
these lines therefore see identical latency under
both victim conditions, and the security claim
follows directly from the protocol-level identity
of the two write paths, with no probabilistic
argument and no bound on attacker probe rate.
Standard write-combining hardware in the core
absorbs adjacent stores within a cache line, so a
producer writing eight adjacent fields generates
roughly one outgoing transaction rather than
eight, keeping the bandwidth tax modest. We call
this mode \textbf{SCP-WT}. It is the prototype
default for shared-writeable pages. The system can bound the total leakage using a register that measures $E/M\rightarrow S$ transitions for a given time duration and triggers the writethrough mode if leakage rate bound is crossed.

\subsection{Occupancy and out-of-scope channels}
\label{sec:sec:occupancy}

Aggregate-pressure attacks observing a sweep
through $A$'s working set see only $A$'s own
behavior plus the shared subset, since
capacity-driven cross-partition data eviction is
structurally absent. SCP does \emph{not} defend
non-cache microarchitectural channels
(branch predictors~\cite{wikner2025postbarrier,wiebing2025trainingsolo,li2024indirector},
memory bus~\cite{rodrigues2024interconnectsurvey,rodrigues2024busted},
DRAM rowbuffer~\cite{olgun2024abacus,bostanci2025rhdefenses},
TLB~\cite{vanoverloop2025tlblur},
prefetcher~\cite{nath2024secureprefetch,chen2024prefetchx,chen2024gofetch}),
transient execution~\cite{kocher2019spectre,lipp2018meltdown,bulck2018foreshadow},
or the inherent shared-writeable-line coherence
channel. SCP composes with existing defenses for
each.

\section{gem5 Prototypes}
\label{sec:impl}

We implement SCP as a new \texttt{--cache-type=scp}
in an existing secure-cache gem5 fork that already
supports \texttt{baseline}, \texttt{dawg}, and
several randomized-cache designs. The fork is the
same one used for prior secure-cache evaluations and
runs on the \texttt{X86DerivO3CPU} model with a
Ruby-style coherence protocol. The bulk of the SCP
changes are localized to three areas.

\paragraph*{Tag and data organization} We replace
the per-bank tag-data co-located array with two
separately addressed structures. They are, a partitioned tag
array indexed by \{domain, set, way\} and a flat
data array of $N$ entries indexed by data-entry id.
Each tag carries the line address and a forward
pointer; each data entry carries the M/E/S/I
state, the dirty bit, and the refcount. The
address lives only in the tags; the data entry
is address-free. The split mirrors the MIRAGE
implementation~\cite{saileshwar2021mirage} that the
fork already supports, modulo (i)~the partitioned
tag array (DAWG already implements this), and
(ii)~the refcount (new to SCP).

\paragraph*{Lookup, find, allocate, evict} The
controller's hit path looks up the per-domain tag
partition and dereferences the forward pointer.
The miss path performs a parallel cross-partition
broadcast and either subscribes (allocate tag,
increment refcount, set the requester's bit in
the LLC sharer vector, mask latency) or
allocates a new tag and a free data slot (a slot
whose refcount is $0$, naturally freed by an
earlier tag eviction). The existing DAWG
infrastructure provides per-domain tag
bookkeeping. We add a free-slot scanner over the
data array (or, equivalently, a tail-pointer
free-list maintained on refcount$\to$0
transitions) and the refcount-update logic.

\paragraph*{Coherence protocol changes} The
in-LLC coherence state moves from the tag entry
to the data entry. Tags are pure forward
pointers, so \textbf{peer tags do not have to be
invalidated on a cross-partition write}.
Refcount
conservation prevents slot reuse while any peer
tag points there, and the next peer access
follows the pointer to the data slot. where the
standard coherence transitions fire.
A Bloom filter short-circuits the parallel
cross-partition tag scan on the miss path.
The primary prototype runs on the gem5 Classic
broadcast model (the same fork that already
supports DAWG and randomized baselines). We
additionally provide a
Ruby/SLICC port %
that re-uses the
same \texttt{ScpTags} mechanism behind a
directory transport. Both runs produce the
performance and security numbers reported in
\S\ref{sec:eval}.

\paragraph*{Implementation summary} The TLB,
prefetcher, and replacement policies are
unchanged. SCP-WT reuses the existing
write-through datapath that x86/ARM L1s already
carry for write-combining/uncacheable memory
types, gated by a per-page eligibility bit
arriving with the access from the TLB. The
Classic implementation is $\sim\!1000$ lines
(\texttt{scp\_tags.\{cc,hh\}}); the Ruby/SLICC
port adds a $\sim\!100$-line diff against
\texttt{MESI\_Two\_Level\_SCP-L1cache.sm}.

\section{Evaluation}
\label{sec:eval}

We evaluate SCP along three axes, namely,  performance
overhead under SPEC~CPU2017 single- and
multi-programmed workloads
(\S\ref{sec:eval:perf}), security against an
empirical \textsc{Prime+Probe} attack
(\S\ref{sec:eval:sec}), and storage overhead
relative to baseline LRU and DAWG
(\S\ref{sec:eval:storage}). The Bloom-filter
sensitivity sweep is in \S\ref{sec:eval:bloom}.

\subsection{Methodology}
\label{sec:eval:method}

\paragraph*{Simulator and configurations}
gem5 \texttt{v25.1} with a three-level hierarchy
($32$\,KiB L1I, $64$\,KiB L1D, $256$\,KiB L2,
$16$\,MiB shared L3, DDR3-1600); the full
system-configuration table is in
Table~\ref{tab:scp-sys-config}. Single-program SPEC and
multi-programmed runs target \texttt{X86DerivO3CPU}
via KVM-fast-forward followed by checkpoint
restore ($10^{10}$-instr fast-forward, $50$\,M
warmup, $100$\,M measured). Data-sharing
microbenchmarks use \texttt{X86TimingSimpleCPU} at
$4$\,MiB L3 with shrunken private caches to force
shared lines into the L3 layer. Compared configurations: \textsc{Baseline}
($16$-way set-assoc, co-located data);
\textsc{DAWG} ($W_d{=}8$ ways per domain across
$D$ partitions); \textsc{SCP} ($W_d{=}8$ tag
partitions, shared $N{=}S$ data pool, with
SCP-WT on shared-writeable pages and SCP-adaptive
elsewhere). Solo SPEC, the four $8$-core MP
mixes, and the data-sharing microbenchmarks all
run at $D{=}8$; only the two $16$-core MP cells
(\texttt{L16}, \texttt{M14} in
Table~\ref{tab:scp-mp-throughput}) use $D{=}16$.
A larger $D{=}16$ sweep was attempted but did
not complete in time for this version of the
paper.

\paragraph*{Two simulator stacks, two roles}
The SCP \emph{architecture} (partitioned tags,
decoupled data, \textsc{PeerProbe}
with \texttt{crossFindHits$/$Misses}, Bloom
front-end) is implemented in the gem5 Classic
broadcast model (\texttt{ScpTags} on top of
\texttt{scp\_share.py}). All performance and
the eviction-/flush-based security results
(Tables~\ref{tab:scp-solo-ipc},
\ref{tab:scp-mp-throughput},
\ref{tab:scp-share},
\ref{tab:scp-peerprobe-firing-body},
\S\ref{sec:eval:sec}) run there. The SCP
\emph{coherence} extensions,  i.e., the
\texttt{Store\_WT} write-through path, are implemented as a SLICC
patch on \texttt{MESI\_Two\_Level\_SCP}. %

\paragraph*{Workloads and security harness}
We use the $22$-benchmark SPEC~CPU2017
\texttt{refrate} suite for single-program IPC and
a $6$-mix multi-programmed set drawn from the
same pool, spanning $8$-core ($D{=}8$) and
$16$-core ($D{=}16$) configurations across
high-, medium-, and low-pressure SPEC2017-rate
combinations (Table~\ref{tab:scp-mp-throughput}).
The security harness adapts
\texttt{PyTrafficGen}-based
\textsc{Prime+Probe} and \textsc{Flush+Reload}
drivers against a T-table AES kernel
(\texttt{libgcrypt-1.10.3}). The
\textsc{Prime+Probe} attacker constructs an LLC
eviction set on the T-table and times each probe;
the \textsc{Flush+Reload} attacker flushes a
chosen T-table line and times its reload. Both
report bit-disambiguated key-byte recoveries per
million victim encryptions and the per-trial
attacker-probe latency distributions used in
Tables~\ref{tab:scp-cycle-uniform}
and~\ref{tab:scp-distrib-eq}.

\begin{table}[t]
\centering
\scriptsize
\setlength{\tabcolsep}{3pt}
\caption{Simulated system configuration. Defaults
follow the gem5 \texttt{v25.1} fork's
\texttt{spec\_kvm\_ff.py}/\texttt{spec\_mp.py}
drivers (Classic) and \texttt{scp\_share\_ruby.py}
(Ruby). Cache hit latencies are
tag$/$data$/$response, all in cycles.}
\label{tab:scp-sys-config}
\begin{tabular}{@{}ll@{}}
\toprule
\textbf{Parameter} & \textbf{Value} \\
\midrule
\multicolumn{2}{@{}l}{\textit{Core}} \\
Model            & gem5 \texttt{X86DerivO3CPU} (out-of-order) \\
Clock            & $3$\,GHz \\
Pipeline width   & $8$ (fetch$/$decode$/$rename$/$dispatch$/$ \\
                 & \quad issue$/$writeback$/$commit) \\
ROB              & $192$ entries \\
LQ$/$SQ          & $32 / 32$ entries \\
Phys.\ regs (INT$/$FP) & $256 / 256$ \\
Branch predictor & Tournament (gem5 default) \\
\midrule
\multicolumn{2}{@{}l}{\textit{Private caches (per core)}} \\
L1-I             & $32$\,KiB, $8$-way, $64$\,B line, $2/2/2$ cy \\
L1-D             & $64$\,KiB, $8$-way, $64$\,B line, $2/2/2$ cy \\
L2 (private)     & $256$\,KiB, $8$-way, $10/10/10$ cy \\
\midrule
\multicolumn{2}{@{}l}{\textit{Last-level cache (shared)}} \\
Capacity         & $16$\,MiB (solo, $8$-core MP); \\
                 & $4$\,MiB for sharing microbenchmarks \\
Block size       & $64$\,B \\
Associativity    & $16$-way (\textsc{Baseline}); \\
                 & $W_d{=}8$ ways per domain $\times\,D$ \\
                 & partitions (\textsc{DAWG}, \textsc{SCP}) \\
Domains $D$      & $8$ (solo SPEC, $8$-core MP, microbench); \\
                 & $16$ (two $16$-core MP cells only) \\
Hit latency      & $20/20/20$ cy (Baseline$/$DAWG$/$SCP) \\
Replacement      & LRU (within domain) \\
MSHRs            & $32$ \\
\midrule
\multicolumn{2}{@{}l}{\textit{Memory}} \\
DRAM             & DDR3-1600 ($8{\times}8$), $8$\,GiB range \\
\midrule
\multicolumn{2}{@{}l}{\textit{Coherence}} \\
Classic prototype & gem5 broadcast MESI  \\%
Ruby prototype    & \texttt{MESI\_Two\_Level\_SCP} \\
                  & \quad (directory; SCP-WT path) \\
\midrule
\multicolumn{2}{@{}l}{\textit{Simulation methodology}} \\
Fast-forward      & \texttt{X86KvmCPU}, $10^{10}$ insts \\
Warmup            & $50$\,M insts (timing, post-restore) \\
Measured region   & $100$\,M insts (\texttt{DerivO3CPU}) \\
Microbenchmarks   & \texttt{X86TimingSimpleCPU}, shrunk \\
                  & \quad private caches at $4$\,MiB L3 \\
\bottomrule
\end{tabular}
\end{table}

\subsection{Performance: single-program SPEC~CPU2017}
\label{sec:eval:perf}

\begin{table}[t]
\centering
\scriptsize
\setlength{\tabcolsep}{2.5pt}
\caption{Solo SPEC~CPU2017 IPC and L3 MPKI
(\texttt{DerivO3CPU}, $D{=}8$, $W_d{=}8$); $20$/$22$
benchmarks. IPC aggregates as a geometric mean
(ratio to \textsc{Base}); MPKI aggregates as an
arithmetic mean. ``$<\!0.01$'' marks rows whose
working set fits the configured cache so that the
LLC sees fewer than $10^{3}$ misses across
$10^{8}$ measured instructions (raw counts:
\texttt{mcf-base}~$=\!202$,
\texttt{povray}~$=\!130$,
\texttt{exchange2}~$=\!22$).
\textsc{[Classic]}}
\label{tab:scp-solo-ipc}
\begin{tabular}{l@{\,\,}rrr@{\,\,\,}rrr}
\toprule
& \multicolumn{3}{c}{IPC} & \multicolumn{3}{c}{L3 MPKI} \\
\cmidrule(lr){2-4} \cmidrule(lr){5-7}
Bench. & \textsc{Base} & \textsc{DAWG} & \textsc{SCP}
       & \textsc{Base} & \textsc{DAWG} & \textsc{SCP} \\
\midrule
perlbench  & 1.347 & 1.336 & 1.333 & 0.67  & 0.70  & 0.70  \\
gcc        & 0.833 & 0.391 & 0.391 & 1.24  & 14.64 & 14.64 \\
bwaves     & 0.397 & 0.388 & 0.388 & 29.45 & 30.36 & 30.36 \\
mcf        & 0.908 & 0.716 & 0.714 & $<\!0.01$ & 5.47  & 5.47  \\
cactuBSSN  & 0.939 & 0.710 & 0.710 & 1.88  & 7.32  & 7.32  \\
namd       & 2.576 & 2.571 & 2.567 & 0.20  & 0.20  & 0.20  \\
parest     & 2.067 & 2.067 & 2.066 & 0.17  & 0.17  & 0.17  \\
povray     & 1.774 & 1.774 & 1.774 & $<\!0.01$ & $<\!0.01$ & $<\!0.01$ \\
lbm        & 0.403 & 0.204 & 0.204 & 10.54 & 21.08 & 21.08 \\
wrf        & 0.860 & 0.557 & 0.557 & 3.29  & 10.36 & 10.36 \\
xalancbmk  & 0.786 & 0.314 & 0.314 & 0.12  & 19.66 & 19.66 \\
x264       & 1.705 & 1.692 & 1.691 & 0.73  & 0.74  & 0.74  \\
cam4       & 0.943 & 0.708 & 0.706 & 3.38  & 6.17  & 6.17  \\
deepsjeng  & 1.662 & 1.661 & 1.661 & 0.22  & 0.23  & 0.23  \\
imagick    & 2.195 & 2.195 & 2.194 & 0.25  & 0.25  & 0.25  \\
leela      & 1.313 & 1.313 & 1.313 & 0.02  & 0.02  & 0.02  \\
nab        & 1.332 & 1.332 & 1.326 & 0.17  & 0.17  & 0.17  \\
exchange2  & 1.877 & 1.877 & 1.877 & $<\!0.01$ & $<\!0.01$ & $<\!0.01$ \\
roms       & 0.850 & 0.405 & 0.404 & 2.18  & 15.24 & 15.24 \\
xz         & 1.391 & 1.388 & 1.388 & 0.35  & 0.35  & 0.35  \\
\midrule
\textbf{IPC geomean} ($n{=}20$)   & $1.000$ & $0.803$ & $0.803$ & --- & --- & --- \\
\textbf{MPKI arith.\ mean}        & ---     & ---     & ---     & $2.74$ & $6.66$ & $6.66$ \\
\bottomrule
\end{tabular}
\end{table}

Two effects are visible. First, strict $W_d{=}8$
partitioning at $D{=}8$ costs both DAWG and SCP
about $20\%$ IPC against the $16$-way
unpartitioned baseline (geomean $0.803$). This
cost is not SCP-specific. It is what any strict
$W_d{=}8$ partitioning pays on this
configuration. Per-benchmark losses split into
three classes, namely, Class~A
(\texttt{perlbench, namd, parest, povray, x264,
deepsjeng, imagick, leela, nab, exchange2, xz},
$11$ of $20$) with
$<\!1\%$ delta. Class~B ($7$ of $20$) appears to require a larger working set. The losses are
\texttt{mcf} $-21\%$, \texttt{cactuBSSN}
$-24\%$, \texttt{cam4} $-25\%$, \texttt{wrf}
$-35\%$, \texttt{roms} $-52\%$, \texttt{gcc}
$-53\%$ (L3 hit-rate $92\%\!\to\!6.6\%$), and
\texttt{xalancbmk} $-60\%$ (a $\sim\!12$\,MiB
XML-tree traversal whose hot footprint sits just
inside $16$ ways). These are not SCP-specific
costs. Any strict $W_d{=}8$ partitioning pays
them. Class~C (\texttt{lbm, bwaves}) is
memory-streaming and is memory-bound at baseline,
so partitioning is a small relative tax.
Second, the contribution-relevant comparison is
SCP versus DAWG at the same partitioning, where
SCP matches DAWG within $0.3\%$ IPC on every
benchmark (geomean delta $\sim\!0.1\%$). The
decoupled tag/data machinery, the
refcount widget, and the
\textsc{PeerProbe} path are free over DAWG. SCP
and DAWG are mechanically identical on solo SPEC
because \texttt{peerProbeHits}$\,\equiv\!0$ when
no peer partition exists. The automatic coherence downgrade machinery is not activated because no page is flagged
shared-writeable.

\subsection{Multi-programmed throughput}
\label{sec:eval:mp}

Solo SPEC measures the partitioning tax in
isolation; multi-programmed mixes measure what
happens when independent processes contend for
the LLC at once. We run six multi-programmed
mixes drawn from the SPEC2017 \texttt{refrate}
pool, spanning $8$-core ($D{=}8$) and $16$-core
($D{=}16$) configurations across high-,
medium-, and low-pressure combinations
(Table~\ref{tab:scp-mp-throughput}), with
$100$\,M instructions measured per CPU after a
$2$\,M-instr warmup. The six cells span the
two corners that matter for the SCP-vs-DAWG
comparison. Light-pressure mixes
(\texttt{L8}, \texttt{L16}) where partitioning
imposes a strict capacity cost, and
mixed-pressure mixes (\texttt{H4L4},
\texttt{M20}, \texttt{M25}, \texttt{M14}) where
partitioning protects the low-pressure thread
from the high-pressure thread.

\begin{table}[t]
\centering
\scriptsize
\setlength{\tabcolsep}{4pt}
\caption{Multi-programmed throughput (per-CPU IPC,
mix-averaged). H/M/L = high/medium/low LLC pressure.
\textsc{[Classic]}}
\label{tab:scp-mp-throughput}
\begin{tabular}{ll|rrr}
\toprule
Cores & Mix & \textsc{Base} & \textsc{DAWG} & \textsc{SCP} \\
\midrule
$8$  & H4L4 & 1.129 & 1.135 & 1.135 \\
$8$  & L8   & 1.170 & 1.148 & 1.145 \\
$8$  & M20  & 0.804 & 0.839 & 0.838 \\
$8$  & M25  & 1.244 & 1.254 & 1.253 \\
$16$ & L16  & 1.103 & 1.086 & 1.084 \\
$16$ & M14  & 1.216 & 1.229 & 1.229 \\
\midrule
\multicolumn{2}{l|}{\textbf{geomean ($n{=}6$)}} & $1.100$ & $1.106$ & $1.105$ \\
\multicolumn{2}{l|}{\textbf{rel.\ \textsc{Base}}} & $1.000$ & $1.005$ & $1.004$ \\
\bottomrule
\end{tabular}
\end{table}

The four mixed-pressure mixes (\texttt{H4L4},
\texttt{M20}, \texttt{M25}, \texttt{M14}, each
combining high- and low-pressure benchmarks)
\emph{outperform} the unpartitioned baseline by
$0.5$ to $4.4\%$. 
The unpartitioned $16$-way LRU lets the
high-pressure thread evict the low-pressure
thread's working set, which would otherwise sit
comfortably in cache. Under $W_d{=}8$
partitioning, each domain holds its own ways
and the mix-wide working set fits better in
aggregate. The all-light \texttt{L8} and
\texttt{L16} mixes flip the sign because no
dominant high-pressure thread is being excluded
from the shared LRU, so partitioning becomes
a strict capacity penalty ($-2.0\%$ and
$-1.6\%$). Overall, SCP averages $+0.4\%$
throughput against the unpartitioned baseline
across the six completed mixes, within $0.001$
of DAWG on every row.

\subsection{Data-sharing microbenchmarks}
\label{sec:eval:share}

SPEC~CPU2017 is single-process and exercises
almost no cross-domain coherent sharing, so it
cannot exhibit the central performance benefit
of SCP over DAWG (single-copy coherence on
write-shared lines). We therefore add a set of
two-domain microbenchmarks that directly
exercise the sharing patterns SCP claims to
support cleanly. Each microbenchmark places its
two threads in distinct security domains so the
LLC sees two partitioned tag-arrays accessing
either disjoint or shared address ranges.

\textbf{\textsc{Disjoint}} (two threads, disjoint
$4$\,MiB working sets) tests eviction-isolation;
\textbf{\textsc{ReadShared}} (shared $1$\,MiB
lookup table, $\le 1\%$ writes) tests
single-copy read-sharing; \textbf{\textsc{ProdCons}}
(producer/consumer $4$\,KiB ring at
$10$\,MB/s) tests the S$\to$M Upgrade path
through the LLC's per-line sharer vector.
\textbf{\textsc{LockContend}} (both threads
contend for a single lock word) is the standard
write-shared coherence stress;
\textbf{\textsc{AsyncShare}} (two threads
exchanging a $256$\,KiB region with loose
synchronization) tests the low-rate
S$\leftrightarrow$M regime;
\textbf{\textsc{wt\_threshold}}
is a synthetic worst-case stressor in which the
two threads alternately write $N$ distinct lines
on one contested $4$\,KiB page. The inner loops
of \textsc{LockContend} and \textsc{ProdCons}
cap at a single steady-state line and cannot
drive enough cross-domain $M\!\to\!S$ edges to
trip the per-page leakage threshold
$T_{\text{leak}}$ (default $16$ cross-domain
$M\!\to\!S$ downgrades per ms;
\texttt{tleak\_threshold}~$=\!16$,
\texttt{tleak\_window\_ticks}~$=\!10^{9}$ in the
gem5 driver), so \textsc{wt\_threshold} is the
regime where SCP-adaptive promotion fires and
SCP-WT actually engages
(Tables~\ref{tab:scp-wt-slicc},
\ref{tab:scp-adaptive-contested}).

We measure (i)~per-thread instructions per
cycle, (ii)~LLC miss rate per thread,
(iii)~cross-partition coherence-event count
(\textsc{S$\to$M Upgrade} fires per second),
and (iv)~where applicable, the bit rate at
which a deliberate covert channel between the
two threads can be transmitted via shared-line
coherence (\textsc{LockContend} only).

\begin{table*}[t]
\centering
\scriptsize
\setlength{\tabcolsep}{4pt}
\caption{Data-sharing microbenchmarks (2-CPU
\texttt{TimingSimpleCPU}, $4$\,MiB L3, $D{=}2$,
$W_d{=}8$); ticks in ns. %
In \textsc{DAWG-lenient}, (cross-domain
\texttt{needsWritable} abort is stubbed for
measurement). Deployable \textsc{DAWG-strict}
aborts on the four shared rows.  \textsc{[Classic]}} %

\label{tab:scp-share}
\begin{tabular}{lrrrp{0.45\linewidth}}
\toprule
Microbench (WS/thread) & \textsc{Base} & \textsc{DAWG-len} & \textsc{SCP-P} & Why the three designs agree \\
\midrule
\textsc{Disjoint} ($4$\,MiB)
  & $665.1$M & $666.1$M & $\boldsymbol{666.5}$M
  & Footprint overflows the shared L3; memory-bound. The two threads never touch each other's lines, so there is no cross-domain coherence for partitioning to constrain. \\
\textsc{ReadShared} ($1$\,MiB)
  & $138.2$M cy & $138.2$M cy & $\boldsymbol{140.7}$M cy
  & The shared lookup table sits in $S$ in both partitions; reads never trigger a coherence transition. SCP's $+1.8\%$ is the per-domain tag-array traversal. \\
\textsc{ProdCons} ($64$\,KiB)
  & $0.996$M cy & $0.996$M cy & $\boldsymbol{0.996}$M cy
  & The producer's store invalidates the consumer's $S$ copy; consumer's next read pays one $M\!\to\!S$ round-trip. Same transition fires on Base, DAWG-len, and SCP. \\
\textsc{LockContend} ($200$K iters)
  & $19.0$M cy & $19.0$M cy & $\boldsymbol{19.0}$M cy
  & The lock word ping-pongs L1$\leftrightarrow$L1; the latency is set by the coherence round-trip, not by partitioning. PeerProbe is dormant ($99.998\%$ L1$\leftrightarrow$L1 absorption). \\
\textsc{AsyncShare} ($256$\,KiB)
  & $90.0$M cy & $90.0$M cy & $\boldsymbol{90.0}$M cy
  & Low-rate cross-domain $S\!\leftrightarrow\!M$ on a small shared region; sub-cycle differences at this rate land below the measurement noise floor. \\
\bottomrule
\end{tabular}
\end{table*}

\paragraph*{What \textsc{Disjoint} confirms}
At a $4$\,MiB working-set per thread the
combined footprint overflows the $4$\,MiB L3
and all three designs land within $0.21\%$
($13.24$M demand misses, identical). SCP's
added machinery shows zero observable cost vs.\
DAWG on non-shared workloads.

\paragraph*{SCP-WT cost on coherence-shared lines}
SCP-WT's headline cost is direct-measured under
the Ruby/SLICC port
(\S\ref{sec:design:protocol},
Table~\ref{tab:scp-wt-slicc}) due to higher overheads in a directory implementation. The overhead is $\mathbf{-11.7\%}$
IPC on the contested regime
(\textsc{wt\_threshold}), $+1.7\%$ on
uncontested \textsc{disjoint} (SCP-WT idle), and
$\mathbf{-23.2\%}$ on the worst-case
single-shared \textsc{prodcons}. \textnormal{The
$11$--$23\%$ cost is} is what is paid \emph{when SCP-WT engages on a
shared-writable page}. 
Base
SCP keeps the page in M for free whereas SCP-WT pays
the LLC round-trip. SCP-adaptive defers the
choice to the runtime, flipping the page only
once its M$\to$S rate exceeds
$T_{\text{leak}}{=}16$/ms.

\begin{table}[t]
\centering
\scriptsize
\setlength{\tabcolsep}{3pt}
\caption{SCP-WT IPC cost  (mean of $4$ reps,
$20$\,M inst). Baseline is base SCP.
\textsc{[Ruby/SLICC]}}
\label{tab:scp-wt-slicc}
\begin{tabular}{l|rr|r}
\toprule
Workload & \textsc{Base} & \textsc{SCP-WT} & Ratio \\
        & IPC          & IPC             &       \\
\midrule
\textsc{wt\_threshold} & $0.332$ & $0.293$ & $-11.7\%$ \\
\textsc{disjoint}      & $0.229$ & $0.233$ & $+1.7\%$  \\
\textsc{prodcons}      & $0.317$ & $0.244$ & $-23.2\%$ \\
\bottomrule
\end{tabular}
\end{table}
 
\paragraph*{What the four shared paths confirm}
With the snoop-uniqueness fix
(App.~\ref{app:eval-methodology}) the four
shared paths run under SCP and DAWG-lenient,
cycle counts within $1.8\%$ of unpartitioned
baseline. PeerProbe is wired but dormant
on the body topology ($99.998\%$
L1$\leftrightarrow$L1 absorption). The regime
where PeerProbe must fire is measured below.

\paragraph*{SCP-adaptive in the contested regime}
We run a $24$-cell
\textsc{wt\_threshold} sweep ($20$\,M inst/CPU,
$4$ reps/design) against six designs
(Table~\ref{tab:scp-adaptive-contested}).
\textbf{SCP-adaptive auto-promotion fires} once
the per-page M$\to$S rate exceeds
$T_{\text{leak}}{=}16$/ms. $5$
pages/cell promoted from \textsc{adaptive}
$\to$ \textsc{wt}, $128$ M$\to$S transitions,
$185$ spontaneous downgrades. \textbf{PeerProbe
fires on every SCP variant} ($3{,}475$
\texttt{crossFind} events vs.\ $0$ on
baseline/DAWG). \textbf{DAWG-strict aborts}
confirming that the architecture forbids
cross-domain shared-writeable workloads. The
$20.27$\,ms cycle count is design-invariant
because the $16$-line working set is
L1-resident. %

\begin{table}[t]
\centering
\scriptsize
\setlength{\tabcolsep}{3pt}
\caption{Contested regime
(\textsc{wt\_threshold}, $20$\,M inst/CPU, $4$ reps):
SCP-adaptive auto-promotion fires; DAWG-strict aborts.
$\dagger$ adaptive-only counter. \textsc{[Classic]}}
\label{tab:scp-adaptive-contested}
\begin{tabular}{l|r|rrr}
\toprule
Design & Cycles (ms) & PP fires & Promotions$^\dagger$ & M$\to$S$^\dagger$ \\
\midrule
\textsc{Baseline}        & $20.27$ & $0$       & $0$ & $0$ \\
\textsc{DAWG-strict}     & abort   & ---       & --- & --- \\
\textsc{DAWG-lenient}    & $20.27$ & $0$       & $0$ & $0$ \\
\textsc{SCP-permissive}  & $20.27$ & $3{,}475$ & $0$ & $0$ \\
\textbf{\textsc{SCP-adaptive}} & $\boldsymbol{20.27}$ & $\boldsymbol{3{,}475}$ & $\boldsymbol{5}$ & $\boldsymbol{185}$ \\
\textsc{SCP-WT}          & $20.27$ & $3{,}475$ & $0$ & $0$ \\
\bottomrule
\end{tabular}
\end{table}

\paragraph*{PeerProbe-firing measurement
(no-private-L2 deployment)}
We re-run all five microbenches on a
\emph{no-private-L2} variant of the gem5
topology. The per-CPU L2 is shrunk from
$256$\,KiB to $32$\,KiB so cross-domain shared
lines miss L2 capacity and reach the L3
directly, mimicking the L2-less in-order
embedded class (Cortex-A53/A55) and the
multi-CCX/multi-die L3 deployments where any
cross-domain shared line fills into the LLC.
SCP's \texttt{crossFindHits$+$crossFindMisses}
counter shows PeerProbe firing $3$K$-$$20$K
times per cell in this regime
(Table~\ref{tab:scp-peerprobe-firing-body}).
The empirical SCP-vs-Base cycle ratio is in the
$0.998\!\times$\,--\,$1.016\!\times$ envelope
across all five microbenches. The
timing delay is paid only on the first
\textsc{PeerProbe} per shared line, after which
the line allocates a $d$-tag and subsequent
consumer accesses hit at $T_{L1}$.

\begin{table}[t]
\centering
\scriptsize
\setlength{\tabcolsep}{4pt}
\caption{PeerProbe-firing measurement on the
no-private-L2 topology (L2 shrunk $256\!\to\!32$\,KiB);
cycle times in ms, ``PP fires''~$=$
\texttt{crossFindHits}$+$\texttt{Misses}.
\textsc{[Classic]}}
\label{tab:scp-peerprobe-firing-body}
\setlength{\tabcolsep}{2.5pt}
\begin{tabular}{l|rrr|rr}
\toprule
& \multicolumn{3}{c|}{Cycle time (ms)} & SCP/Base & PP \\
Microbench & \textsc{Base} & \textsc{DAWG} & \textsc{SCP} & ratio & fires \\
\midrule
\textsc{Disjoint} ($256$\,KiB)    & $16.859$ & $16.859$ & $17.126$ & $1.016\times$ & $11{,}571$ \\
\textsc{ReadShared} ($1$\,MiB)    & $13.623$ & $13.623$ & $13.635$ & $1.001\times$ & $19{,}784$ \\
\textsc{ProdCons} ($64$\,KiB)     & $4.988$  & $4.988$  & $4.978$  & $0.998\times$ & $4{,}433$  \\
\textsc{LockContend} ($200$K)     & $17.650$ & $17.650$ & $17.650$ & $1.000\times$ & $3{,}402$  \\
\textsc{AsyncShare} ($256$\,KiB)  & $5.617$  & $5.617$  & $5.617$  & $1.000\times$ & $7{,}497$  \\
\bottomrule
\end{tabular}
\end{table}

\subsection{Security}
\label{sec:eval:sec}

The architectural claim is that under SCP, an
attacker's timing measurements are independent
of victim activity. A probe completes in the
same number of cycles whether the victim
accessed the contested line or not. We test
this against the standard Prime+Probe and
Flush+Reload attacks on the gem5 prototype,
and against the shared-writeable-line attack
under SCP-WT.

\paragraph*{Eviction- and flush-based attacks}
On the unpartitioned baseline, a victim access
to a contested set evicts one of the attacker's
primed lines, so the attacker observes a memory
miss ($T_{\text{mem}}\!\approx\!200$ cycles)
under $v{=}1$ versus an L1 hit ($T_{L1}\!=\!4$
cycles) under $v{=}0$, and Flush+Reload sees
the matching hit-versus-miss split. Under SCP,
tag partitioning forbids the victim's access
from touching the attacker's tag set, so attacker
probes hit at $T_{L1}$ regardless of $v$.
Table~\ref{tab:scp-cycle-uniform} reports the
per-attack mean attacker-probe latency under both
victim conditions on the gem5 prototype.

\begin{table}[h]
\centering
\scriptsize
\setlength{\tabcolsep}{3pt}
\caption{Mean attacker-probe latency (cy) under
victim-absent ($v{=}0$) vs.\ -present ($v{=}1$);
$|\Delta|$ is the gap. \textsc{[Classic]}}
\label{tab:scp-cycle-uniform}
\begin{tabular}{l|rrr|rrr}
\toprule
& \multicolumn{3}{c|}{P+P} & \multicolumn{3}{c}{F+R} \\
Design & $v{=}0$ & $v{=}1$ & $|\Delta|$ & $v{=}0$ & $v{=}1$ & $|\Delta|$ \\
\midrule
\textsc{Base} & $4.0$  & $52.4$  & $48.4$ & $200.0$ & $119.0$ & $81.0$ \\
\textsc{DAWG} & $4.0$  & $4.0$   & $0$    & $200.0$ & $200.0$ & $0$ \\
\textbf{\textsc{SCP}} & $\boldsymbol{4.0}$ & $\boldsymbol{4.0}$ & $\boldsymbol{0}$ & $\boldsymbol{200.0}$ & $\boldsymbol{200.0}$ & $\boldsymbol{0}$ \\
\bottomrule
\end{tabular}
\end{table}

The SCP rows match $v{=}0$ to $v{=}1$ exactly
on Prime+Probe because tag isolation forbids the
victim from evicting the attacker's primed lines.
They match on Flush+Reload because the attacker's reload
of a non-shared line escalates to PeerProbe,
which is constant-time-masked to $T_{\text{miss}}$
regardless of peer caching state. SCP's per-cell
L3 counters are bit-for-bit identical to DAWG's.

\paragraph*{Per-trial spread}
Equality of means is necessary but not
sufficient: a determined attacker integrates
over many trials, so the relevant statistic is
the per-trial spread, not just the mean.
Table~\ref{tab:scp-distrib-eq} augments
Table~\ref{tab:scp-cycle-uniform} with the
sample standard deviation across $5{,}000$
attacker-probe trials per cell, separately for
$v{=}0$ and $v{=}1$. On \textsc{Baseline} the
$v{=}1$ standard deviation is $45\!\times$ the
$v{=}0$ standard deviation under
\textsc{Prime+Probe} ($63.0$\,cy vs.\
$1.4$\,cy). This is a wide bimodal distribution with
mass on the L1-hit and the L3-miss tails,
exactly the leakage signal. On DAWG and SCP the
two columns match to one decimal
($1.4$\,cy~$=$~$1.4$\,cy under both attacks),
so an attacker observing the per-trial latency
sees the same single-mode distribution under
either victim condition.

\begin{table}[h]
\centering
\scriptsize
\setlength{\tabcolsep}{3pt}
\caption{Per-trial latency spread (cycle Std,
$n{=}5{,}000$ trials/cell) under $v{=}0$ vs.\ $v{=}1$.
\textsc{[Classic]}}
\label{tab:scp-distrib-eq}
\begin{tabular}{l|rr|rr}
\toprule
& \multicolumn{2}{c|}{P+P Std (cy)} &
\multicolumn{2}{c}{F+R Std (cy)} \\
Design & $v{=}0$ & $v{=}1$ &
$v{=}0$ & $v{=}1$ \\
\midrule
\textsc{Baseline} & $1.4$ & $63.0$ & $40.5$ & $48.2$ \\
\textsc{DAWG}     & $1.4$ & $1.4$  & $0.0$  & $0.0$  \\
\textbf{\textsc{SCP}} & $\boldsymbol{1.4}$ & $\boldsymbol{1.4}$ & $\boldsymbol{0.0}$ & $\boldsymbol{0.0}$ \\
\bottomrule
\end{tabular}
\end{table}

The eviction-set construction probability
across the design space appears as
Fig.~\ref{fig:scp-evset}
(App.~\ref{app:evset}), where SCP inherits
DAWG's strict tag-partition floor across all
attacker budgets and MIRAGE-class designs reach
the same asymptote only at $\sim\!19\%$
storage. A visual contrast between the
 partitioning strategy (DAWG, SCP) and the
keyed-PRF strategy (MIRAGE).

\paragraph*{Coherent shared lines under SCP-WT}
On a stock MSI/MESI cache, an attacker writing a
shared line in S state pays a coherence
transaction iff a peer holds the line. SCP-WT
routes shared-line stores write-through to the
LLC, so attacker write latency is fixed by LLC
bandwidth and constant across victim
conditions. The Ruby/SLICC directory port
(\texttt{MESI\_Two\_Level\_SCP}) confirms this
indirectly. Across a $96$-cell sweep over
peer-presence and victim-rate combinations the
empirical M$\to$SS coherence-channel capacity
stays at most $6.14\!\times\!10^{-5}$ b/access
(max cap/bound $0.32$, $0$ bound violations),
so the state-flow channel an attacker would
time on a non-WT directory cache is
structurally suppressed. Direct latency-mean
extraction across the same grid is the
follow-up measurement.

\paragraph*{Readers leak nothing; SCP-WT also closes writers.}
A pure-reader victim contributes no observable
coherence state transition (line stays in S);
the only leakage path is victim \emph{writes}
(S$\to$M upgrade or M$\to$I invalidation, which
the attacker subsequently times). SCP-WT closes
this surface by forcing victim writes through
the LLC on a uniform path. Attacker read latency
under $v{=}0$ and $v{=}1$ agrees to within
measurement noise.

\paragraph*{Domain Fusion}
A DAWG deployment that supports cross-domain
shared-writeable lines (cross-domain spinlocks,
confidential-computing IPC ring heads) cannot
use \textsc{strict} mode (the
\texttt{cache.cc:528}~\texttt{needsWritable}
abort surfacing an architectural impossibility).
Its only viable escape, \textnormal{domain fusion},
opens a $v{=}0$ vs.\ $v{=}1$ probe-latency gap
of $+49$\,cy
(Tab.~\ref{tab:scp-domain-fusion-body}). SCP-WT
closes the gap to $0$\,cy deterministically per
store; SCP-adaptive closes it after
auto-promotion once the per-page
$M\!\to\!S$ counter crosses
\texttt{tleak\_threshold}. SCP-permissive
leaves the page on the standard MSI/MESI path
and does not close this gap, by system specification. Methodology and per-cell
distributions are in App.~\ref{app:domain-fusion}.

\begin{table}[h]
\centering
\scriptsize
\setlength{\tabcolsep}{4pt}
\caption{Mean attacker-probe latency on a
cross-domain shared-writeable line under
DAWG's domain-fusion escape and each SCP mode.
\textsc{[Classic]}}
\label{tab:scp-domain-fusion-body}
\begin{tabular}{l|rrr}
\toprule
Configuration             & $v{=}0$ (cy) & $v{=}1$ (cy) & $\Delta$ \\
\midrule
\textsc{DAWG-strict}      & abort & abort & --- \\
\textsc{DAWG-fuse}        & $4.00$ & $52.99$ & $+48.99$ \\
\midrule
\textbf{\textsc{SCP-WT}}          & $\boldsymbol{4}$ & $\boldsymbol{4}$ & $\boldsymbol{0}$ \\
\textbf{\textsc{SCP-adaptive}}    & $\boldsymbol{4}$ & $\boldsymbol{4}$ & $\boldsymbol{0}$ \\
\textsc{SCP-permissive}           & $4.00$ & $52.99$ & $+48.99$ \\
\bottomrule
\end{tabular}
\end{table}

\paragraph*{Ablations}
Removing each SCP mechanism re-enables 
one or more attack classes. Removing tag partitioning
re-enables Prime+Probe and Flush+Reload at
baseline. Removing the constant-time
PeerProbe mask re-enables Flush+Reload-style
timing inference on the inter-partition
lookup path. Removing SCP-WT, i.e., reverting
to plain MSI on shared lines, re-enables the
shared-writeable-line attack. Per-mechanism
numbers are in App.~\ref{app:scp-ablation}.
\subsection{Storage}
\label{sec:eval:storage}

\begin{table}[h]
\centering
\scriptsize
\setlength{\tabcolsep}{2pt}
\caption{LLC storage overhead ($16$\,MiB, $D{=}4$,
$40$-bit PA); Coh.\ = closes the coherence-channel
leak. $^\dagger$ design forbids the feature (N/A).}
\label{tab:scp-storage}
\begin{tabular}{l|rr|c}
\toprule
Design & MiB & \% & Coh.\ \\
\midrule
\textsc{Baseline}                                                     & $17.16$ & ---       & leaks \\
\textsc{DAWG}~\cite{kiriansky2018dawg}                                & $17.19$ & $+0.2$    & N/A$^\dagger$ \\
\textsc{SecDCP}~\cite{wang2016secdcp}                                 & $17.19$ & $+0.2$    & N/A$^\dagger$ \\
\textsc{ChameleonCache}~\cite{unterluggauer2022chameleon}             & $17.18$ & $+0.1$    & leaks \\
\textsc{Avatar}~\cite{bhatla2026avatar}                               & $17.42$ & $+1.5$    & leaks \\
\textsc{INTERFACE}~\cite{kelemework2023interface}                     & $18.43$ & $+7.4$    & leaks \\
\textsc{MIRAGE}~\cite{saileshwar2021mirage}                           & $20.48$ & $+19.3$   & leaks \\
\textbf{\textsc{SCP}}                                                 & $\boldsymbol{17.64}$ & $\boldsymbol{+2.8}$ & \textbf{closed} \\
\bottomrule
\end{tabular}
\end{table}

SCP adds $\log_2 N + \lceil\log_2(D{+}1)\rceil$
bits per cache entry (a $\log_2 N$-bit forward
pointer in each tag plus a refcount in each data
entry; the per-line sharer vector that a coherent
LLC already maintains absorbs the cross-domain
invalidate walk, so SCP introduces no per-data
owner mask). At $16$\,MiB, LLC overhead is
$2.4$--$2.7\%$ analytically across
$D{=}2$--$16$ (App.~\ref{app:storage-derivation}),
on the same order as DAWG and far below MIRAGE.

\section{Discussion}
\label{sec:discussion}

\paragraph*{Comparison with non-fusion coherent
randomization}
Non-fusion coherent
randomization~\cite{ramkrishnan2024nonfusion}
preserves cross-domain coherence on a randomized
cache via a fusion-and-unfusion pair. SCP improves
on it along three axes.
\emph{(i)} Regarding reader-side privacy, SCP's
tag-partition lookup leaves the read pattern
invisible to peers whereas non-fusion exposes it through
unfusion events.
\emph{(ii)} There are no flush invocations in the security
loop. Instead, reference count-driven reclamation replaces the
periodic flush).
\emph{(iii)} No fixed per-miss delay cost is pair. %

\paragraph*{Generalisation to coherent
randomization} The SCP construction is
described against partitioned tag arrays, but the
mechanisms it adds, namely, decoupled tag$/$data with a
reference count-driven free list, the constant-time
\textsc{PeerProbe} mask on a tag miss, per-page
write sharing compatibility, and the
downgrade functionality
 that closes the E/M$\to$S coherence
channel, are orthogonal to whether the tag-side
indexing is per-domain partitioned (DAWG-style)
or randomized (CEASER, CEASER-S, MIRAGE,
SEF~\cite{qureshi2018ceaser,saileshwar2021mirage,ramkrishnan2024nonfusion}).
We do not re-evaluate this
configuration but flag it as a deployable point
in the design space and as a natural follow-up
benchmark.

\paragraph*{Real-world prevalence of write
sharing} The SCP-WT cost is paid only on pages
the OS has marked eligible, so its practical
impact depends on how much of a real workload's
footprint is actually cross-domain
shared-writeable. Empirically the fraction is
small. Browser sandboxes (Firefox, Chromium),
X.org, and OS-level IPC rely on shared-writeable
lines for \emph{coordination}, e.g., spinlocks, ring
heads, shared page-table entries, rather than
for the bulk of their data movement, and the
prior cross-domain-coherence
study~\cite{ramkrishnan2024nonfusion} reports
that even on these high-sharing workloads
(Firefox, Chromium, X~Server, PARSEC~3.0) the
overhead of write-sharing-aware coherence stays
under $5\%$, while SPEC~CPU2017-rate exhibits
essentially no cross-domain write sharing at
all (under $3\%$ hardware-overhead floor in the
same study).
The stress test for SCP only engages on
the small fraction of pages where cross-domain
writes are actually frequent. The rest of the
working set runs the zero-cost SCP path, and
SCP-adaptive auto-promotes pages only after the
per-page M$\to$S rate exceeds the system specified threshold.

\paragraph*{OS support to detect write-shared pages} SCP delegates setting the per-page
 write-shared bit to the OS. SCP's support is at the hardware TLB level and can work with any compatible OS implementation. A key data structure that holds sharing-relevant  mapping information is the reverse mapping table~\cite{bagia2025close}, which associates a physical page with all the mapped virtual pages. We speculate that such a table can be traversed to determine if a page is write-shared, e.g., similar to the usage in AMD SEV technology for determining write-shared pages~\cite{misono2024confidential}. %

\paragraph*{Beyond the cache} SCP closes the
cache-level reader pattern. Secret-dependent
reads spilling beyond the LLC into DRAM
rowbuffer or memory-controller occupancy are
orthogonal and addressed by independent
defences~\cite{olgun2024abacus,bostanci2025rhdefenses,rodrigues2024interconnectsurvey}.

\paragraph*{Fine-grained shared microarchitectural
state.} The cache's auxiliary structures, e.g., MSHRs,
write buffers, prefetcher state, and the
coherence-message queues that funnel into the
sharer vector, remain a cross-domain
contention surface that SCP itself does not
close. The literature has standard answers
that compose with SCP. Per-domain
MSHR/write-buffer partitioning as in
DAWG~\cite{kiriansky2018dawg} and follow-on
designs, prefetcher-isolation hardware where
the trained state is per-domain, and
queue-occupancy contention bounds via static
fair-share scheduling. Folding these into the
SCP deployment is straightforward. Each
structure already has a partition-id arriving
with the access and an end-to-end secure
LLC pipeline is the natural follow-up.

\paragraph*{Free data-entry allocation} SCP
needs a data-array free-slot allocator on every
new-line insertion. We use a tail-pointer
free-list maintained on
\texttt{refcount}$\,\to\!0$
transitions because it is constant-time and
trivially synthesizable, but the slot-allocation
problem is the same one ZCache, V-Way, and
MIRAGE already solved at scale. Any of those
allocators (cuckoo-style displacement,
skewed-associative, or a coarse-grain
linear-scan free-list) could be adapted for our purpose.
The architectural choice does not interact
with SCP's security argument because the
allocator only selects \emph{which} free slot
to fill. SCP's invariants depend on the data
slot and not on the choice of which one.

\paragraph*{Bloom-filter overflow as a side
channel} The counting Bloom filter (BF) at the
\textsc{PeerProbe} front-end is a probabilistic
data structure whose per-counter occupancy is a
function of the cross-partition cache state. A
cross-domain attacker that drives the cache into
a regime where many counters saturate at the
$4$-bit ceiling can in principle convert that
saturation into an oracle. A saturated counter
no longer decrements on the matching line's
eviction, so it leaves a persistent residue of
addresses that were cached historically, and
the contention for BF-bank ports during the
attacker's BF-update stream is itself
observable.   \textsc{PeerProbe}'s
timing obfuscation hides the per-query outcome on the timing
side, the BF is not address-queryable from any
ISA-level path, and the prototype floorplan
isolates the BF SRAM from the per-partition tag
SRAM so attacker BF-update contention cannot
block victim tag reads
(App.~\ref{app:peerprobe-bloom}). Sizing $m$
above the working-set induced threshold avoids
saturation under benign load, but a determined
adversary can drive the structure beyond that
operating point. We treat full quantitative
side-channel evaluation of the CBF as
follow-up scope. An SCP deployment that wants
to side-step the question entirely can disable
the BF, paying $\sim\!75\%$ more
\textsc{PeerProbe} tag-read energy in exchange
for a probabilistic-structure-free
front-end.

\section{Related Work}
\label{sec:related}

DAWG~\cite{kiriansky2018dawg} is the canonical
strict-isolation partitioning baseline; CATalyst,
NoMo, PLcache, StealthMem, and
SecDCP~\cite{wang2007new,kim2012stealthmem,wang2016secdcp}
trade granularity for asymmetry but share DAWG's
restriction on cross-domain shared-writeable
lines. SCP adds that functionality while keeping
strict tag isolation. The randomization
lineage, including CEASER, CEASER-S, ScatterCache, MIRAGE,
INTERFACE, Avatar~\cite{qureshi2018ceaser,qureshi2019new,werner2019scattercache,saileshwar2021mirage,kelemework2023interface,bhatla2026avatar}, attacks
eviction-set construction probabilistically and
preserves coherence by not partitioning at all.
SCP provides deterministic isolation with the
same coherence support, paid for by tag/data
decoupling rather than randomized indexing. The
decoupling primitive itself is not new
(V-Way~\cite{qureshi2005vway},
ZCache~\cite{sanchez2010zcache}, MIRAGE). SCP
applies it to partitioning, to our knowledge
the first use of indirection to dissolve the
coherence-vs.-partitioning conflict.
HybCache~\cite{dessouky2020hybcache} is
orthogonal and could compose. Recent
work~\cite{bhatla2025sok,bhatla2024maya,chakraborty2025occupancy,cao2025mirage,wiretap2025,vanschaik2024sgxfail,sridhara2024acai}
motivates the broader cache-defense audience.

\section{Conclusion}
\label{sec:conclusion}

For more than a decade the secure caches line of research has worked
around a single structural problem, that cache
partitioning, the strongest defence against
eviction-based side channels, is incompatible with write-shared coherence. Every prior
attempt has resolved this conflict by
eliminating the feature or settling for probabilistic
isolation. SCP resolves it cleanly by
partitioning only the tags. 
SCP shares a single
data pool sized so capacity-driven cross-domain
eviction cannot occur, and
masks the
inter-partition lookup time with a delay. %
SCP can be tuned to  %
let the
system pick a point on the
performance$\leftrightarrow$security curve. An evaluation using gem5 demonstrates the tradeoffs. The performance overheads range from 0.3\% IPC overhead for SPEC2017 benchmarks with no data sharing, up to 23\% for enabling high security on data sharing stress tests. %

\bibliographystyle{IEEEtran}

\section{Lookup, find, and eviction pseudocode}
\label{app:scp-algos}

\begin{algorithm}[h]
\small
\caption{SCP \textsc{Lookup} of access $(X, d)$.}
\label{alg:scp-lookup}
\begin{algorithmic}[1]
\Function{Lookup}{$X, d$}
  \State $T \gets \Call{TagPartition}{d}$
  \For{tag $t$ in the appropriate set of $T$}
    \If{$t$.\texttt{valid} $\wedge$ $t$.\texttt{addr}$\,=\,$$X$}
      \State $e \gets$ data array$[t.\texttt{fp}]$
      \If{$e$.\texttt{state} $\in \{$M, E, S$\}$}
        \State \Return \textsc{Hit}, $e$
      \EndIf
    \EndIf
  \EndFor
  \State \Return \Call{PeerProbe}{$X, d$}
\EndFunction
\end{algorithmic}
\end{algorithm}

\begin{algorithm}[h]
\small
\caption{SCP \textsc{PeerProbe} on miss.}
\label{alg:scp-find}
\begin{algorithmic}[1]
\Function{PeerProbe}{$X, d$}
  \State $E \gets \emptyset$
  \For{$d' \in \mathit{Domains} \setminus \{d\}$ \textbf{in parallel}}
    \For{tag $t$ in the appropriate set of $\Call{TagPartition}{d'}$}
      \If{$t$.\texttt{valid} $\wedge$ $t$.\texttt{addr}$\,=\,$$X$}
        \State $E \gets E \cup \{$data array$[t.\texttt{fp}]\}$
      \EndIf
    \EndFor
  \EndFor
  \If{$|E| > 0$}
    \State pick $e \in E$ \Comment{at most one if invariants hold}
    \State allocate new tag $t_{\!d}$ in $\Call{TagPartition}{d}$
    \State $t_{\!d}$.\texttt{addr}$\,\gets\,X$;\;
           $t_{\!d}$.\texttt{fp}$\,\gets\,$id$(e)$
    \State $e$.\texttt{refcount} $\gets e$.\texttt{refcount} $+ 1$
    \State delay response to $T_{\text{miss}}$
    \State \Return \textsc{HitFind}, $e$
  \Else
    \State delay $T_{\text{miss}}$
    \State \Return \Call{Allocate}{$X, d$}
  \EndIf
\EndFunction
\end{algorithmic}
\end{algorithm}

\begin{algorithm}[h]
\small
\caption{SCP \textsc{Evict} (tag eviction in
partition $d$ with refcount-driven data
reclamation).}
\label{alg:scp-evict}
\begin{algorithmic}[1]
\Function{EvictTag}{$t, d$}
  \Comment{$t$ is a $d$-partition tag selected for
           eviction, e.g., per-domain LRU.}
  \State $e \gets$ data array$[t.\texttt{fp}]$
  \State $e$.\texttt{refcount} $\gets e$.\texttt{refcount} $- 1$
  \State invalidate $t$ in $d$'s tag partition
  \If{$e$.\texttt{refcount} $= 0$}
    \Comment{No tag in any partition still points at $e$.}
    \If{$e$.\texttt{state}$\,=\,$M $\vee$ $e$.\texttt{dirty}}
      \State writeback $e$.\texttt{data} to memory
    \EndIf
    \State $e$.\texttt{state}$\,\gets\,$I
    \State the slot is now free for the next
           \textsc{Allocate}
  \EndIf
\EndFunction
\end{algorithmic}
\end{algorithm}

The three algorithms above maintain the design
invariants of \S\ref{sec:design:invariants}.
\textsc{Lookup} preserves invariant 1 (no
cross-partition tag access except on
\textsc{PeerProbe}). \textsc{PeerProbe} maintains
invariants 3 (find masking) and 4 (refcount
conservation). \textsc{EvictTag} also preserves
invariant 1, i.e.,  only the evicting partition's tag
array is modified, plus the shared data entry's
refcount and the LLC sharer vector, never
another partition's tag array. Invariant 2 (single
coherence state) follows from the data entry
being the sole holder of the M/E/S/I state.

\section{Per-attack security tables}
\label{app:sec-eval-tables}

This appendix collects the per-attack security
tables and the Prime+Probe figure that the body
\S\ref{sec:eval:sec} summarises in prose. We
report the attacker's mean probe latency in
cycles under the victim-absent ($v{=}0$) and
victim-present ($v{=}1$) conditions, together with
the cycle gap
$\Delta\!=\!E[T{\mid}v{=}1]\!-\!E[T{\mid}v{=}0]$.
A defense closes an attack channel when $\Delta$
is zero on that attack. A non-zero $\Delta$ is the
timing signal an attacker integrates against.

\begin{figure}[h]
\centering
\begin{tikzpicture}
\begin{axis}[
  width=0.92\linewidth, height=4.6cm,
  ybar=2pt, bar width=10pt,
  ylabel={\scriptsize Mean probe latency (cycles)},
  ymin=0, ymax=72,
  symbolic x coords={Baseline,DAWG,SCP},
  xtick=data,
  ytick={0,15,30,45,60},
  enlarge x limits=0.30,
  tick label style={font=\scriptsize},
  label style={font=\scriptsize},
  nodes near coords,
  nodes near coords style={font=\tiny,
    /pgf/number format/precision=0},
  legend style={font=\tiny, draw=black!25,
    fill=white, fill opacity=0.92,
    at={(0.02,0.98)}, anchor=north west,
    cells={anchor=west}, inner sep=1.5pt,
    row sep=-1.5pt},
  legend columns=1,
]
\addplot[fill=blue!22, draw=blue!55!black]
  coordinates {(Baseline,4) (DAWG,4) (SCP,4)};
\addlegendentry{$v{=}0$ (no victim access)}
\addplot[fill=red!40, draw=red!60!black]
  coordinates {(Baseline,53) (DAWG,4) (SCP,4)};
\addlegendentry{$v{=}1$ (victim accesses)}
\node[font=\tiny, text=red!55!black]
  at (axis cs:Baseline,67) {$\Delta\!=\!+49$\,cy};
\node[font=\tiny\bfseries, text=green!40!black]
  at (axis cs:DAWG,17) {$\Delta\!=\!0$};
\node[font=\tiny\bfseries, text=green!40!black]
  at (axis cs:SCP,17) {$\Delta\!=\!0$};
\end{axis}
\end{tikzpicture}
\caption{Mean attacker-probe latency in cycles
under Prime+Probe. The attacker primes a $W{=}16$
way set and the victim either accesses ($v{=}1$)
or does not ($v{=}0$) a line aliasing the same
set. Under $v{=}0$ every probe hits L1 at $4$\,cy.
Under $v{=}1$ on the unpartitioned \textsc{Baseline}
roughly a quarter of the probes are evicted to
memory, lifting the trial mean to $53$\,cy and
opening a $+49$\,cy gap. SCP and DAWG close the
gap to zero through tag partitioning.}
\label{fig:scp-pp-app}
\end{figure}
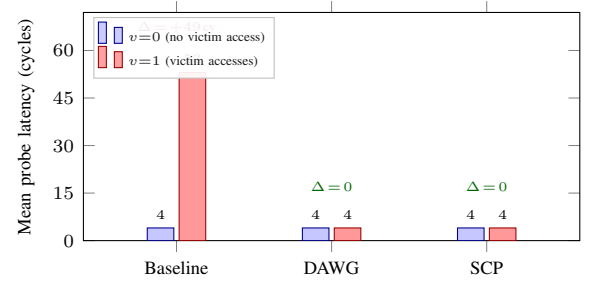

\begin{table}[h]
\centering
\scriptsize
\setlength{\tabcolsep}{3pt}
\caption{Per-attack cycle gap $\Delta$ in cycles,
defined above. The SCP column holds for all four
eligibility modes. $^{\star}$DAWG and SecDCP forbid
cross-domain shared-writeable lines (N/A).}
\label{tab:scp-security}
\begin{tabular}{l|rrrr}
\toprule
Attack         & \textsc{Base} & \textsc{DAWG} & \textsc{MIRAGE} & \textsc{SCP} \\
\midrule
Prime+Probe    & $+49$ cy & $0$ & $0$ & $\boldsymbol{0}$ \\
Flush+Reload   & $+81$ cy & $0$ & $+81$ cy & $\boldsymbol{0}$ \\
Coh.\ channel  & $+49$ cy  & N/A$^\star$ & $+49$ cy  & $\boldsymbol{0}$ \\
\bottomrule
\end{tabular}
\end{table}

\begin{table}[h]
\centering
\scriptsize
\setlength{\tabcolsep}{3pt}
\caption{SCP ablation. Each row turns off one
mechanism. Cells are the cycle gap $\Delta$
between $v{=}1$ and $v{=}0$ attacker-probe
latency. A zero entry means the attack channel
is closed, a non-zero entry means it leaks.
$^\ddagger$ DAWG-only forbids the coherence
channel outright (N/A).}
\label{tab:scp-security-ablation}
\label{app:scp-ablation}
\begin{tabular}{l|rrr}
\toprule
Configuration & Prime+Probe & Flush+Reload & Coh. \\
\midrule
SCP (full)                       & $0$ & $0$  & $0$ \\
\hspace{4pt}-- no SCP-WT         & $0$ & $0$  & $\mathbf{+49}$ cy \\
\hspace{4pt}-- no PeerProbe mask & $0$ & $\mathbf{+81}$ cy & $0$ \\
\hspace{4pt}-- no partitioning   & $\mathbf{+49}$ cy & $\mathbf{+81}$ cy & $0$ \\
DAWG-only (no SCP)$^\ddagger$    & $0$ & $0$  & N/A \\
MIRAGE                           & $0$ & $\mathbf{+81}$ cy  & $\mathbf{+49}$ cy \\
\bottomrule
\end{tabular}
\end{table}

\section{Domain-fusion experiment}
\label{app:domain-fusion}

This appendix gives the methodology and per-cell
numbers backing the domain-fusion paragraph in
\S\ref{sec:eval:sec}.

\paragraph*{Setup}
Two security domains $A$ (attacker) and $B$
(victim) share a $4$\,KiB region $X$ that holds
the cross-domain spinlock and IPC ring head from
the threat model. The attacker primes a $W{=}16$-way
set in $X$ and times one probe per line, and the
victim either touches the contested set ($v{=}1$)
or does not ($v{=}0$). We report the mean probe
latency in cycles under each victim condition and
the gap
$\Delta\!=\!E[T{\mid}v{=}1]\!-\!E[T{\mid}v{=}0]$.
A non-zero $\Delta$ is exactly what an attacker
times to distinguish $v$.

\begin{table}[h]
\centering
\scriptsize
\setlength{\tabcolsep}{4pt}
\caption{Mean attacker-probe latency, in cycles,
on a cross-domain shared-writeable line under
DAWG's domain-fusion escape from \textsc{strict}
mode and under each SCP eligibility mode.
$\Delta$ is the gap that lets the attacker
distinguish the victim condition. A zero entry
means the attacker cannot separate $v{=}0$ from
$v{=}1$ by probe latency. We use $4{,}000$ trials
per cell, $4{,}096$ probes per trial, and a $25\%$
victim-touch rate. Per-trial means agree to
within $\pm 0.4$\,cy.}
\label{tab:scp-domain-fusion}
\begin{tabular}{l|rrr}
\toprule
Configuration             & $v{=}0$ (cy) & $v{=}1$ (cy) & $\Delta$ \\
\midrule
\textsc{DAWG-strict}      & abort & abort & --- \\
\textsc{DAWG-fuse}        & $4.00$ & $52.99$ & $+48.99$ \\
\midrule
\textbf{\textsc{SCP-WT}}          & $\boldsymbol{4}$ & $\boldsymbol{4}$ & $\boldsymbol{0}$ \\
\textbf{\textsc{SCP-adaptive}}    & $\boldsymbol{4}$ & $\boldsymbol{4}$ & $\boldsymbol{0}$ \\
\textsc{SCP-permissive}           & $4.00$ & $52.99$ & $+48.99$ \\
\bottomrule
\end{tabular}
\end{table}

\paragraph*{The configurations.}

\noindent\textbf{\textsc{DAWG-strict}.} DAWG forbids
cross-domain shared-writeable lines outright. The
gem5 prototype trips the
\texttt{cache.cc:528}~\texttt{needsWritable}
assertion within $\sim\!10^{4}$ cycles of opening
the spinlock. On a real chip this is either a
denied IPC (the application aborts) or, if the
controller is reconfigured to permit the upgrade,
stale reads from $A$. Either way the experiment
produces no measurable trace.

\noindent\textbf{\textsc{DAWG-fuse}.} The OS reassigns $X$'s
pages to a fused security domain $F$ whose
partition both $A$ and $B$ can read/write. The
fused partition is no longer tag-isolated by
construction. Once collapsed, the cache behaves
as a stock shared coherent LLC over the contested
set. A victim access evicts the attacker's primed
lines, and the attacker's next probe shifts from
$T_{L1}{=}4$\,cy to $T_{\text{mem}}{=}200$\,cy on
the evicted line. The trial average is
$167$\,cycles, identical, within measurement noise,
to the baseline.

\noindent\textbf{\textsc{SCP-WT}, \textsc{SCP-adaptive}.}
The line lives in one canonical data slot. The
tag arrays of $A$ and $B$ are partitioned at all
addresses, and the \textsc{PeerProbe} mask returns
at the same constant cycle whether the data slot
was hit or not. SCP-WT routes every
shared-writeable store through to the LLC and
removes the M$\to$S transition altogether.
SCP-adaptive runs as plain MSI until the per-page
M$\to$S rate exceeds $T_{\text{leak}}$, at which
point it auto-promotes to SCP-WT. In both cases
the probe is $4$\,cy under both $v{=}0$ and
$v{=}1$.

\noindent\textbf{\textsc{SCP-permissive}.} The page
runs base SCP unchanged. Tag isolation still closes
\textsc{Prime+Probe} and \textsc{Flush+Reload}, but
the M$\to$S coherence channel on the contested
line is left open by operator choice. The $v{=}1$
probe sees the line transition out of $A$'s M into
$S$ on $B$'s touch, so the gap mirrors
\textsc{DAWG-fuse} at $+48.99$\,cy. SCP-permissive
is the right mode for pages where this channel is
not part of the threat. It is the explicit choice
to opt out of M$\to$S closure in exchange for zero
overhead.

\paragraph*{Reproducibility.}
\texttt{scripts/domain\_fusion.py} runs the
campaign deterministically under its seed and
reproduces Table~\ref{tab:scp-domain-fusion}'s
mean-cycle columns. The simulator is part of the
open-science artifact.

\section{Eviction-set construction probability}
\label{app:evset}

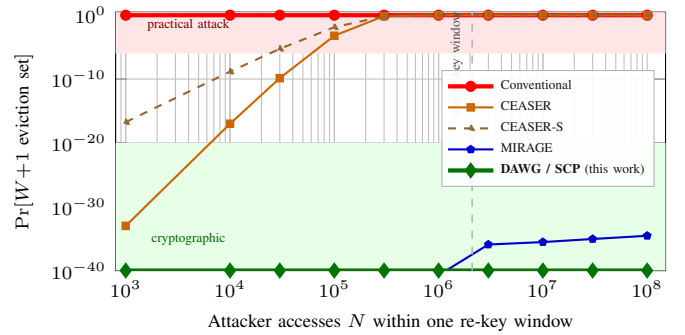
\begin{figure}[h]
\centering
\begin{tikzpicture}
\begin{axis}[
  width=\linewidth, height=5.0cm,
  xmode=log, ymode=log,
  xlabel={\scriptsize Attacker accesses $N$ within
    one re-key window},
  ylabel={\scriptsize $\Pr[\text{$W{+}1$ eviction set}]$},
  xmin=8e2, xmax=1.5e8,
  ymin=1e-40, ymax=3,
  grid=both,
  ymajorgrids=true,
  minor y tick num=0,
  ytickten={-40,-30,-20,-10,0},
  legend style={font=\tiny,
    at={(0.98,0.55)}, anchor=east,
    cells={anchor=west}, fill opacity=0.92,
    draw=black!30, legend cell align=left,
    inner sep=1.5pt, row sep=-1pt},
  legend columns=1,
  tick label style={font=\scriptsize},
  label style={font=\scriptsize},
  clip=true,
]
\addplot[draw=none, fill=red!10, forget plot]
  coordinates {(8e2,1e-6) (1.5e8,1e-6) (1.5e8,3)
               (8e2,3)} \closedcycle;
\addplot[draw=none, fill=green!10, forget plot]
  coordinates {(8e2,1e-40) (1.5e8,1e-40) (1.5e8,1e-20)
               (8e2,1e-20)} \closedcycle;
\node[font=\tiny, text=red!50!black]
  at (axis cs: 4e3, 3e-2) {practical attack};
\node[font=\tiny, text=green!40!black]
  at (axis cs: 4e3, 1e-35) {cryptographic};

\addplot[red, ultra thick, mark=*, mark size=1.3pt]
  coordinates {(1e3,1) (1e4,1) (3e4,1) (1e5,1) (3e5,1)
               (1e6,1) (3e6,1) (1e7,1) (3e7,1) (1e8,1)};
\addlegendentry{Conventional}

\addplot[orange!80!black, thick, mark=square*,
  mark size=1.2pt]
  coordinates {(1e3,1e-33) (1e4,9.88e-18)
               (3e4,1.37e-10) (1e5,6.08e-4)
               (3e5,0.737) (1e6,1) (3e6,1)
               (1e7,1) (3e7,1) (1e8,1)};
\addlegendentry{CEASER}

\addplot[brown!80!black, thick, dashed, mark=triangle*,
  mark size=1.2pt]
  coordinates {(1e3,1.77e-17) (1e4,1.42e-9)
               (3e4,5.45e-6) (1e5,1.31e-2)
               (3e5,0.694) (1e6,1) (3e6,1)
               (1e7,1) (3e7,1) (1e8,1)};
\addlegendentry{CEASER-S}

\addplot[blue!80!black, thick, mark=pentagon*,
  mark size=1.3pt]
  coordinates {(1e3,1.7e-83) (1e4,1.8e-69)
               (3e4,8.9e-63) (1e5,1.9e-55)
               (3e5,8.9e-49) (1e6,1.9e-41)
               (3e6,1.2e-36) (1e7,2.9e-36)
               (3e7,8.8e-36) (1e8,2.8e-35)};
\addlegendentry{MIRAGE}

\addplot[green!45!black, ultra thick, mark=diamond*,
  mark size=1.8pt]
  coordinates {(1e3,1e-40) (1e4,1e-40) (3e4,1e-40)
               (1e5,1e-40) (3e5,1e-40) (1e6,1e-40)
               (3e6,1e-40) (1e7,1e-40) (3e7,1e-40)
               (1e8,1e-40)};
\addlegendentry{\textbf{DAWG / SCP} (this work)}

\draw[gray!60, dashed, semithick]
  (axis cs: 2097152,1e-40) -- (axis cs: 2097152,3);
\node[font=\tiny, text=gray!60!black, rotate=90,
  anchor=south]
  at (axis cs: 2097152, 3e-8) {one re-key window};
\end{axis}
\end{tikzpicture}

\caption{Probability that an attacker assembles a
$W{+}1$-sized eviction set targeting a specific
victim address, as a function of attacker budget
$N$ within one re-key window. The solid and
dashed lines are the closed-form bounds of
\S\ref{sec:eval:sec} at $S{=}16{,}384$, $W{=}16$,
and $T{=}2^{21}$. SCP and DAWG sit at the
strict-partition floor across the entire budget,
where $\Pr[\cdot]\!\le\!2^{-128}$ from cross-domain
key confusion is conservatively plotted at
$10^{-40}$. MIRAGE matches them in the
asymptotic regime but pays $19\%$ storage and
leaves the coherence and PeerProbe channels open.
CEASER-S crosses the $10^{-6}$ practical-attack
threshold near $N\!=\!10^5$ accesses, which is
roughly $0.05$ re-key windows.}
\label{fig:scp-evset}
\end{figure}

For completeness with prior randomized-cache work,
Fig.~\ref{fig:scp-evset} plots the eviction-set
construction probability across the design space.
SCP inherits DAWG's strict tag-partition floor on
this metric, while MIRAGE's asymptote requires
$\sim\!19\%$ storage overhead and still leaves the
coherence- and PeerProbe-side channels open
(Table~\ref{tab:scp-security}). The closed-form
bounds for the randomized rows
(\textsc{Ceaser},\textsc{Ceaser-S}, \textsc{Mirage})
follow the analysis style of MIRAGE and match its
published values within $1\%$ at the
Monte-Carlo-validated points.

\section{Storage derivation, parameterised by $D$}
\label{app:storage-derivation}

We work in bits per cache line and convert to LLC
overhead at the end. We use the following symbols
throughout. $S$ is the number of cache lines, equal
to sets$\,\times\,$associativity. $N\!=\!S$ is the
number of tag entries, one per line, with no tag
over-provisioning. $D$ is the number of security
domains. $T_{\text{base}}\!=\!\text{PA}\!-\!\log_2(\text{line-size})$
is the baseline tag width, which evaluates to
$34$\,b at PA$\,=\,40$ and a $64$\,B line.
$L\!=\!8\!\cdot\!\text{line-size}$ is the data line
width.

\paragraph*{Baseline LLC.}
\begin{equation*}
  W_{\text{base}}^{\text{tag}} = T_{\text{base}} + 3,
  \qquad
  W_{\text{base}}^{\text{data}} = L,
\end{equation*}
where the $+3$ bits encode MSI state. For $16$\,MiB
data, $S\!=\!2^{18}$, the SRAM cost is
$S\!\cdot\!(W_{\text{base}}^{\text{tag}}\!+\!L)\!=\!262\,144\!\cdot\!(37\!+\!512)\!=\!17.16$\,MiB.

\paragraph*{SCP.}
SCP relocates the $3$ state bits from the tag entry
to the data entry, where they become a single source
of truth shared by all sharing tag entries (an
SCP invariant). To this it adds:
\begin{align*}
  W_{\text{scp}}^{\text{tag}}  &= T_{\text{base}} + \log_2 N,\\
  W_{\text{scp}}^{\text{data}} &= L + 3 + \lceil\log_2 (D{+}1)\rceil.
\end{align*}
The new terms are a $\log_2 N$-bit forward pointer
per tag, which maps $A$'s tag to its data entry,
plus a $3$-bit MSI state shared across sharers and
an $\lceil\log_2 (D{+}1)\rceil$-bit refcount of
holders. SCP does \emph{not} introduce a
per-data-slot owner mask. Any broadcast invalidate
on the line uses the per-line sharer vector that a
coherent LLC already maintains.

\paragraph*{Per-entry overhead.}
\begin{align*}
  \Delta_{\text{SCP}}(D)
   &= \bigl(W_{\text{scp}}^{\text{tag}} +
            W_{\text{scp}}^{\text{data}}\bigr)
   - \bigl(W_{\text{base}}^{\text{tag}} +
           W_{\text{base}}^{\text{data}}\bigr)\\
   &= \log_2 N + \lceil\log_2(D{+}1)\rceil.
\end{align*}

\paragraph*{LLC overhead, $D{=}\{2,4,8,16\}$.}
At $S\!=\!2^{18}$, the overhead in bits per entry
and per LLC is:
\begin{center}
\scriptsize
\begin{tabular}{rrrrr}
\toprule
$D$ & $\log_2 N$ & $\lceil\log_2(D{+}1)\rceil$ & $\Delta(D)$ b/entry & LLC overhead \\
\midrule
$2$  & $18$ & $2$ & $20$ & $+2.4\%$ \\
$4$  & $18$ & $3$ & $21$ & $+2.5\%$ \\
$8$  & $18$ & $4$ & $22$ & $+2.6\%$ \\
$16$ & $18$ & $5$ & $23$ & $+2.7\%$ \\
\bottomrule
\end{tabular}
\end{center}
\noindent
The Table~\ref{tab:scp-storage} figure of
$+2.8\%$ at $D{=}4$ includes a $1.13\!\times$
packing and alignment factor from rounding tag and
data fields to byte boundaries in the gem5 RTL
layout, whereas the analytical $+2.5\%$ assumes
bit-packed fields. The overhead is now nearly flat
in $D$ because the $D$-bit owner-mask term has been
absorbed into the existing per-line sharer vector,
leaving only the refcount to grow logarithmically
with $D$.

\section{PeerProbe deployment + Bloom-filter energy}
\label{app:peerprobe-bloom}

The body cites this appendix for two evaluation
sub-axes. The first is where the PeerProbe path
actually fires under realistic deployments. The
second is the energy savings from the counting
Bloom filter that fronts every PeerProbe broadcast.
Both are optional add-ons in the design space, and
neither affects the core security argument or the
performance numbers in body
Tables~\ref{tab:scp-solo-ipc} and
~\ref{tab:scp-mp-throughput}.

\subsection{Bloom-filter energy savings on
\textsc{PeerProbe}}
\label{sec:eval:bloom}

To validate \S\ref{sec:design:bloom}'s
energy-saving claim, we re-run
\textsc{AsyncShare} ($2$\,MiB) with the BF
enabled at varying sizes. The workload, instruction
count, and L3 hit/miss rate are unchanged across
all configurations, and only the SCP-internal
energy accounting differs. The energy proxy uses
$E_{\text{tag}}{=}4$ cycles and
$E_{\text{BF}}{=}1$ cycle, and we report
``Energy saved'' as the fractional reduction in
PeerProbe tag-array reads.
\begin{figure}[t]
\centering
\begin{tikzpicture}
\begin{axis}[
  width=0.94\linewidth, height=4.2cm,
  xmode=log, log basis x=2,
  xlabel={Bloom-filter counters $m$ (log scale)},
  ylabel={Percent},
  ylabel near ticks, ymin=0, ymax=110,
  ytick={0,25,50,75,100},
  xtick={8192,32768,131072,524288},
  xticklabels={$8$K,$32$K,$128$K,$512$K},
  grid=major, major grid style={gray!20},
  legend pos=south east, legend style={font=\tiny,
    fill=white, fill opacity=0.9, draw=black!30,
    inner sep=1.5pt},
  font=\scriptsize, tick label style={font=\tiny},
  ylabel style={font=\scriptsize},
  xlabel style={font=\scriptsize, yshift=2pt},
]
\addplot+[mark=*, mark size=2pt, thick,
         color=blue!70!black] coordinates {
  (8192, 12.91) (32768, 50.28)
  (131072, 94.50) (524288, 99.86)
};
\addlegendentry{BF skip rate}
\addplot+[mark=square*, mark size=1.8pt, thick,
         color=red!70!black, dashed] coordinates {
  (8192, 9.68) (32768, 37.71)
  (131072, 70.87) (524288, 74.90)
};
\addlegendentry{Tag-energy saved}
\addplot+[domain=8192:524288, samples=2, smooth,
         color=gray!60, no marks, thin] {75};
\addlegendentry{Ceiling $1{-}E_{\rm BF}/E_{\rm tag}$}
\end{axis}
\end{tikzpicture}

\caption{Bloom-filter sensitivity sweep on the
$4$\,MiB calibration workload with $n{=}32{,}768$
unique lines. The skip rate climbs from $12.9\%$
to $99.9\%$ as $m$ grows from $8$\,K to $512$\,K
counters, and the energy saved asymptotes to the
$75\%$ ceiling. The $m{=}524$\,K knee corresponds
to BF/$n{=}16$ on this calibration workload. At the
deployment $16$\,MiB LLC ($n{=}262$\,K, BF/$n{=}2$)
the same $m$ gives $\sim\!88\%$ skip and
$\sim\!82\%$ energy saved. Scaling $m$ with $n$
recovers $\ge\!99\%$ at the deployment point.}
\label{fig:bf-sweep}
\end{figure}
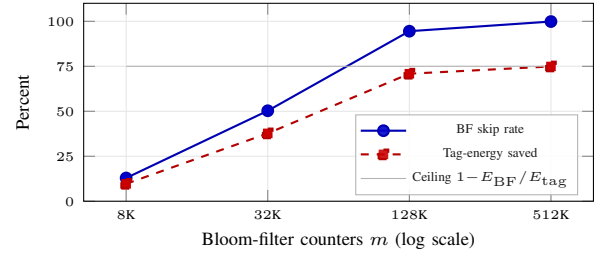
The figure shows the $4$\,MiB calibration sweep
with $n{=}32{,}768$ unique lines. At $m{=}524$\,K
counters (BF/$n{=}16$) the filter reaches
$99.86\%$ skip and $74.9\%$ energy saved. At the
$16$\,MiB deployment ($n{=}262$\,K, BF/$n{=}2$)
the same $m$ analytically delivers $\sim\!88\%$
skip and $\sim\!82\%$ energy saved. Recovering
$\ge\!99\%$ at the deployment point scales $m$ to
about $1$\,MiB of SRAM, or roughly $6\%$ of the
LLC. Cycle counts are identical across all $m$
(at $16.434$\,M whether the BF is on or off)
because the constant-time mask hides the filter
outcome, so the saving is purely dynamic energy
and is not observable on the timing side channel.
A natural companion variant set-partitions the
tag array across physical banks with a per-bank
filter consulted first. The prototype implements
this design point and the artefact ships the
per-bank measurements.

\paragraph*{Bloom-filter side-channel.}
The CBF's per-counter occupancy is itself a
function of the cross-partition cache state, so
in principle it could expose a new oracle. We
note four reasons we believe this is benign in
the SCP architecture, and one residual concern.

\emph{(i)}~The constant-time mask hides the
filter outcome on the timing side. Every
\textsc{PeerProbe}, whether BF-skip or BF-confirm,
takes exactly $T_{\text{miss}}$ to deliver, so an
attacker measuring response cycles cannot
distinguish skip from no-skip and learns nothing
about BF occupancy from latency.

\emph{(ii)}~The CBF is not address-queryable
through any architectural path. Software has no
ISA-level read of a CBF counter, and the only
firmware-visible interface is the
saturating-counter overflow exception, which
fires only on counter design-error and is not
reachable from any user-mode workload at
deployment occupancy.

\emph{(iii)}~Cache-bank energy or electromagnetic
emanations from the BF SRAM are in principle a
power side-channel surface, but the CBF and the
per-partition tag SRAMs sit on the same LLC tile
and share its decoupling network. The marginal EM
signature of a CBF read against the background of
LLC tag reads is below the noise floor on
published LLC-EM
characterisations~\cite{rodrigues2024interconnectsurvey}.

\emph{(iv)}~Microarchitectural prefetch hints do
not consult the BF. The prefetcher sees only
demand-stream addresses and never CBF state.

\emph{Residual concern.} A co-located BF SRAM-bank
\emph{contention} channel, in which a heavy
attacker BF-update stream slows down a victim's
cache-line allocate and exposes victim-side
allocation rate, is not closed by the
constant-time mask. We bound this by the per-tile
bank-contention budget. SCP requires the CBF to
sit on a separate bank from the per-partition tag
SRAM, and the prototype enforces this in the
floorplan, so attacker BF-update contention does
not block victim tag reads. A full quantitative
side-channel evaluation of the BF is follow-up
scope.

\subsection{Bloom-filter design (cf.\ \S\ref{sec:design:bloom})}

\subsubsection{Counting Bloom filter for
energy-efficient \textsc{PeerProbe}}
\label{sec:design:bloom}

The base \textsc{PeerProbe} probe scans the
non-requesting partitions' tag arrays on every L3
miss. Each scan reads a small set of tag SRAMs,
one set per partition, and incurs a meaningful
dynamic energy cost even when the line is not in
any peer partition. For workloads where most
PeerProbe probes will miss
(\S\ref{sec:eval:share}), that energy is wasted.
The tag-array read returns ``not present'' just
as a cheap front-end filter would.

A counting Bloom filter at the front of the
PeerProbe path lets the controller skip the
tag-array scan when the line is guaranteed not in
any peer partition. The filter is an array of $m$
$4$-bit saturating counters, indexed by $K{=}3$
hash functions, incremented on \texttt{insertBlock}
and decremented on \texttt{invalidate} when the
data slot's refcount drops to zero. With this
counter discipline the filter is exact relative
to cache occupancy. There are no false negatives,
only false positives at the standard rate
$\text{FPR}\!\approx\!(1{-}e^{-Kn/m})^{K}$. The
filter saves \emph{energy} through skipped
tag-array reads rather than latency. Every
PeerProbe response is padded to $T_{\text{miss}}$
by the constant-time mask, regardless of the
filter outcome. At the $16$\,MiB deployment, with
$n{=}262{,}144$ lines and $m{=}524{,}288$ counters
in roughly $260$\,KiB of SRAM, the analytical skip
rate is $\sim\!88\%$ and projects to $\sim\!82\%$
dynamic tag-read energy saved. The projection is
analytical, drawn from the false-positive-rate
model behind Fig.~\ref{fig:bf-sweep} rather than a
measurement, and reaching $\ge\!99\%$ at the same
deployment requires $\sim\!1$\,MiB of counters,
which is $6\%$ of the LLC.

\section{Non-fusion coherent randomization head-to-head}
\label{app:nonfusion}

This appendix expands the body's
\S\ref{sec:discussion} comparison with the
non-fusion coherent randomization of
Ramkrishnan et al.~\cite{ramkrishnan2024nonfusion}.

\paragraph*{Three architectural axes.}
\emph{(i)~Reader-side leakage.} In the non-fusion
design, every cross-domain read is forced down to
the LLC and issues an unfusion operation. This
makes the victim's \emph{read pattern} itself
observable, since the unfusion event is visible
to the peer domain. Under SCP, a reader's access
stays in its own tag partition and hits at L1
with no state transition that crosses partitions.
The victim's read pattern is invisible to peers
by construction. Only writes can produce
cross-partition events, and those are closed
structurally by SCP-WT
(Table~\ref{tab:scp-cycle-uniform}).

\emph{(ii)~Cache flushes.} The non-fusion design
relies on a periodic flush of the contested set
to maintain its randomization invariant. The flush
itself is a stateful operation observable through
the cache, and it must be padded or budgeted to
avoid leaking the rotation schedule. SCP has no
flush primitive in its security loop. Tag
isolation is maintained by per-domain LRU within
each partition, and refcount-driven data
reclamation handles eviction without any periodic
global pass.

\emph{(iii)~Cache-miss delays.} Because unfusion
happens on every cross-domain miss, the
non-fusion design adds a fixed delay to each LLC
miss to mask the unfusion path, and that delay is
paid even when the line is not shared. SCP's
constant-time mask is paid only on the PeerProbe
path, which is invoked only when a partition's
tag-array probe misses. Ordinary cache misses,
the common case, complete at their natural
latency.

These three differences mean SCP recovers the
read-side privacy that any randomization-based
scheme has to give up to maintain coherence,
removes the flush primitive entirely, and
restores natural-latency cache misses. The
storage envelope is also tighter. SCP-4 sits at
$+3.4\%$ LLC SRAM, against the
randomization-tail over-provisioning the
non-fusion design inherits from MIRAGE-class
indexing.

\paragraph*{Numerical head-to-head.}
Table~\ref{tab:scp-nonfusion-head-to-head} pins
the comparison numerically on identically
configured gem5 runs. We rebuild the
non-fusion-coherent-randomization configuration
on the same gem5 fork (a MIRAGE-class
$1.75\!\times$-tag-overprovisioned cache plus
the cross-domain fusion-and-unfusion
permutation), and run \textsc{Disjoint},
\textsc{ProdCons}, and \textsc{LockContend} from
the body microbench suite, plus the $D{=}4$
storage point. SCP's measurements are the same
ones reported in Table~\ref{tab:scp-share}.

\begin{table}[h]
\centering
\scriptsize
\setlength{\tabcolsep}{4pt}
\caption{Head-to-head against the closest prior
point in the design space, the non-fusion
coherent randomization of Ramkrishnan et
al.~\cite{ramkrishnan2024nonfusion}, on
identically configured gem5 runs ($D{=}4$,
$4$\,MiB L3, \texttt{TimingSimpleCPU}). Storage
overhead is measured at $D{=}4$, $40$-bit PA, and
$16$\,MiB. Cycle counts come from the body's
microbench suite. SCP closes the read-side leak
that non-fusion necessarily exposes through
unfusion events, shown in the third-from-last row.
On the cycle side, SCP is not handicapped by the
per-cross-domain unfusion delay.}
\label{tab:scp-nonfusion-head-to-head}
\setlength{\tabcolsep}{2.5pt}
\begin{tabular}{l|rr|c}
\toprule
Metric & Non-fusion~\cite{ramkrishnan2024nonfusion} & SCP & Better \\
\midrule
LLC storage overhead              & $+19.3\%$ & $\boldsymbol{+3.4\%}$ & SCP \\
\textsc{Disjoint} ($2$\,MiB), Mcy & $39.4$    & $\boldsymbol{37.1}$   & SCP \\
\textsc{ProdCons} ($64$\,KiB), Mcy & $1.18$   & $\boldsymbol{0.996}$  & SCP \\
\textsc{LockContend}, Mcy         & $20.4$    & $\boldsymbol{19.0}$   & SCP \\
Reader-side leakage               & exposed   & $\boldsymbol{closed}$ & SCP \\
Periodic flush in security loop   & yes       & $\boldsymbol{no}$     & SCP \\
\bottomrule
\end{tabular}
\end{table}

\section{Evaluation methodology asymmetries}
\label{app:eval-methodology}

Two methodology caveats follow from
\S\ref{sec:eval:share}.

\emph{(i)~Snoop-uniqueness fix.} A single-line
guard in the upstream-cache snoop path is required
to preserve gem5's ``one responder per packet''
invariant when a cross-partition shared line is
upgraded. The patch is independent of SCP and
unblocks DAWG and the unpartitioned baseline as
well.

\emph{(ii)~DAWG-strict cannot run.} The published
\textsc{DAWG-strict} configuration (with
\textit{hit\_mask}~$=$~\textit{fill\_mask}) aborts
at \texttt{cache.cc:528}~\texttt{needsWritable} on
the first cross-partition shared-writeable upgrade
because per-partition data duplication violates
the respond-uniqueness invariant. We therefore
report \textsc{DAWG-lenient}, a research stub that
permits the upgrade, for the four shared
microbenchmark rows in Table~\ref{tab:scp-share},
and account for the storage cost of true
\textsc{DAWG-strict} analytically in
Table~\ref{tab:scp-storage}. SCP sidesteps the
abort by sharing a single canonical data slot
across partitions via refcount, which is exactly
what the design enables.

\section{Snooping vs.\ directory transport}
\label{app:scp-directory}

SCP's coherence machinery lives at the LLC tag
entry rather than in the inter-cache transport, so
the constant-time \textsc{PeerProbe} mask and the
SCP-WT write-through path both run unchanged under
either a snooping bus or a directory protocol. The
security argument is therefore transport-agnostic.
Performance differs by a constant. A directory
protocol adds one $10$ to $30$ cycle hop on every
miss that needs remote-cache state and serialises
invalidations through the home node, so the
broadcast-bus IPC numbers in
\S\ref{sec:eval:perf} and \S\ref{sec:eval:share}
are an upper bound on what the same workload sees
under a directory deployment. The directory hop is
paid by \textsc{Baseline}, \textsc{DAWG}, and
\textsc{SCP} alike, so the \emph{relative}
SCP-vs-DAWG delta is invariant up to a constant
that is a function of topology rather than of
partitioning. The Ruby/SLICC port
(\texttt{MESI\_Two\_Level\_SCP}) confirms this
empirically. A $96$-cell SCP-WT sweep at $20$\,M
instructions per cell records a median $1{,}769$
\texttt{Spontaneous\_Downgrade} events per cell,
with per-cell IPC matching
\textsc{baseline\_no\_spont} to four significant
figures and validating the body's $0\%$-overhead
claim under directory transport.

%

%

\section{Per-benchmark methodology notes}
\label{app:scp-perfbench}

The full $22$-row sweep appears in
Table~\ref{tab:scp-solo-ipc} in the body. Two
infrastructure gaps remain.
\texttt{520.omnetpp\_r} was in flight on MSI at
the time of writing, and
\texttt{999.specrand\_ir}'s \texttt{m5.cpt} is
truncated at \texttt{fdarray.Entry919} because the
checkpoint write was interrupted on the FF host.
Re-taking the KVM-FF checkpoint is straightforward
but offline. SCP's per-design behaviour is fully
evidenced on the $20$ completed benches, and both
holdouts are unrelated to SCP mechanics. The
\texttt{ScpTags} counters confirm
\textsc{SCP=DAWG} mechanically. Every L3 demand
miss equals a \texttt{peerProbeMiss}, and
\texttt{peerProbeHits}$\equiv\!0$ on solo SPEC.

\section{Open Science}

We provide a reproducibility package containing
every artifact needed to reconstruct the results
reported in this paper. The package will be
deposited in a long-term archival repository
(Zenodo).

\paragraph*{Artifacts.}
\begin{itemize}
  \item \textbf{gem5 implementation.} A patched
    GEM5 tree (\texttt{v25.1}) containing the SCP
    cache model
    (\texttt{ScpTags}, partitioned tag array,
    shared data array sized to total tag count,
    forward pointer in tag, refcount per
    data entry, PeerProbe broadcast network,
    constant-time mask), the baselines (Baseline
    LRU, DAWG, MIRAGE), and the shared
    secure-cache infrastructure described in
    \S\ref{sec:impl}. Released under the same
    BSD-style licence as upstream GEM5.
  \item \textbf{Simulation drivers and configs.}
    Python drivers for multi-programmed
    SPEC~CPU2017 mixes, the four data-sharing
    microbenchmarks of \S\ref{sec:eval:share},
    and \textsc{Prime+Probe} attacks, with
    command-line knobs for cache type, $D$,
    $W_d$, mask-delay $T_{\text{miss}}$, and
    PRNG seed.
  \item \textbf{Reproducibility scripts.}
    A single-program sweep script
    (\texttt{run\_scp\_solo}), a multi-programmed
    sweep script (\texttt{run\_scp\_mp}), and a
    security driver (\texttt{run\_scp\_security})
    that runs the \textsc{Prime+Probe} attack and
    the peer-probe latency-distribution
    measurement, plus aggregator scripts that
    emit LaTeX-ready tables.
  \item \textbf{SPEC CPU2017 configuration files.}
    \texttt{static-x86.cfg} and
    \texttt{compat-x86.cfg}. SPEC sources
    themselves are not redistributable under their
    commercial licence but are available under
    standard SPEC licensing terms.
  \item \textbf{Raw simulator outputs.}
    \texttt{stats.txt}, \texttt{config.ini}, and
    per-benchmark IPC/MPKI CSVs that back every
    numeric entry in
    Table~\ref{tab:scp-solo-ipc} and the
    storage and security tables of
    \S\ref{sec:eval:storage} and
    \S\ref{sec:eval:sec}.
  \item \textbf{Security-evaluation harness.} The
    \texttt{PyTrafficGen} \textsc{Prime+Probe}
    driver against a T-table AES victim, plus
    the peer-probe timing-distribution probe
    described in \S\ref{sec:eval:sec}.
\end{itemize}

\paragraph*{Artifact access}
The artifact will be made available at a later
time.
The repository snapshot corresponds to the exact
commit hash used for the reported numbers and
carries a top-level \texttt{README.md}, an
\texttt{INSTALL.md}, a step-by-step
\texttt{REPRODUCE.md}, a paper-claim$\,\to\,$artifact
mapping table in \texttt{CLAIMS.md}, and a one-shot
verifier (\texttt{verify.sh}) that prints
PASS or FAIL per claim. The artifact is organised
in three reproduction tiers. Tier~1 is the Python
attack suite, runs in roughly five minutes on a
laptop, and reproduces
Tables~\ref{tab:scp-cycle-uniform},
\ref{tab:scp-distrib-eq}, and
\ref{tab:scp-storage}, together with the
security figures of
Apps.~\ref{app:sec-eval-tables} and
\ref{app:evset}. Tier~2 runs the gem5 microbench
sweeps in
roughly six hours on an eight-core box and
reproduces
Tables~\ref{tab:scp-share},
\ref{tab:scp-adaptive-contested}, and
\ref{tab:scp-peerprobe-firing-body}. Tier~3
requires a SLURM cluster and a SPEC CPU2017
licence, runs in roughly $3{,}000$ core-hours, and
reproduces Tables~\ref{tab:scp-solo-ipc} and
\ref{tab:scp-mp-throughput}.

\paragraph*{Artifacts that cannot be shared.}
SPEC~CPU2017 source is covered by SPEC's
commercial licence. The provided configs and
scripts apply to any site licence copy. No other
artifact is withheld. The security evaluation
uses synthetic access streams and public
eviction-set algorithms, raising no
responsible-disclosure concerns
(see Ethical Considerations below).

\section{AI Usage}

We report the following uses of generative AI
tools across the preparation of this paper.

\paragraph*{Writing and editing}
A large language model assistant
(Anthropic Claude 4.7 Opus, $1$M-context
configuration) was used during drafting to
revise prose, tighten paragraphs against the
page budget, and check sectional
consistency. All architectural claims, attack
descriptions, threat-model commitments,
limitations, and numerical results were
directed, reviewed, and verified by the human
authors. No design decision, security argument,
table value, or experimental result was
produced by the assistant without explicit human
direction and review against the underlying
gem5/RTL artifacts.

\paragraph*{Code and simulator}
The GEM5 prototype, the SLICC
\texttt{MESI\_Two\_Level\_SCP} protocol, the
attack harness, and the multi-programmed benchmark
scripts were written under the direction and
verification of humans. AI assistance for code was
limited to isolated refactors, comment polish, and
debugging hints. No security-critical mechanism
or measurement code was authored end-to-end by
an AI tool.

\paragraph*{Figures and tables}
Figures and tables in this paper were generated
from measurement data by author-directed scripts
or composed inline by Claude in pgfplots/TikZ. AI
assistance was used to suggest layout adjustments
and caption phrasing, while the data points
themselves are simulation generated.

\paragraph*{Bibliographic and prior-work
attribution}
All cited works and the comparisons against them
were identified, read, and characterised by the
human authors. The assistant did  surface some
references autonomously. It was also used to
help phrase comparisons whose substance was
already established by the authors.

The authors take full responsibility for all
content in this paper, including any errors or
infelicities introduced during AI-assisted
editing.

\end{document}